\pdfoutput=1

\documentclass[11pt,twoside,a4paper,cmspaper,final,collab]{cms-tdr}
\def\svnVersion{177391}\def\cmsCernNo{2012-079}\def\cmsCernDate{\today}\def\cmsMessage{Submitted to the Journal of High Energy Physics}

\begin{document}\cmsNoteHeader{QCD-10-029}

\hyphenation{had-ron-i-za-tion}
\hyphenation{cal-or-i-me-ter}
\hyphenation{de-vices}
\RCS$Revision: 123522 $
\RCS$HeadURL: svn+ssh://svn.cern.ch/reps/tdr2/papers/QCD-10-029/trunk/QCD-10-029.tex $
\RCS$Id: QCD-10-029.tex 123522 2012-05-22 19:28:03Z bhatti $
\newcommand\rs{\raisebox{1.0ex}[-1.0ex]}
\newcommand{\ra}{\ensuremath{\rightarrow}}
\newcommand{\znunu}{\ensuremath{{\text Z} \ra \nu\bar{\nu}}}
\newcommand{\zmumu}{\ensuremath{{\text Z} \ra \mu\mu}}
\newcommand{\wmunu}{\ensuremath{{\text W} \ra \mu\nu}}
\newcommand{\wtaunu}{\ensuremath{{\text W} \ra \tau\nu}}
\newcommand{\dphi}{\ensuremath{\Delta \phi}}
\newcommand{\dphijj}{\ensuremath{\Delta \phi_{ j1,j2}}}
\newcommand{\pT}{\ensuremath{{p_{\text T}}\xspace}}
\newcommand{\pts}{\ensuremath{p_{\text T}{\text s}}\xspace}
\newcommand{\Et}{\ensuremath{{E_{\text T}}\xspace}}
\newcommand{\ptjf}{\ensuremath{p_{\rm T}^{ {\rm j}_1} }}
\newcommand{\ptjs}{\ensuremath{p_{\rm T}^{ {\rm j}_2} }}
\newcommand{\ptjt}{\ensuremath{p_{\rm T}^{ {\rm j}_3} }}
\newcommand{\etajf}{\ensuremath{\eta^{ {\rm j}_1} }}
\newcommand{\etajs}{\ensuremath{\eta^{ {\rm j}_2} }}
\newcommand{\etajt}{\ensuremath{\eta^{ {\rm j}_3} }}
\newcommand{\ttj}{\ensuremath{\rm{t}\bar{\rm{t}} + jets}\xspace}
\newcommand{\wj}{\ensuremath{\rm W + jets}\xspace}
\newcommand{\zj}{\ensuremath{\rm Z + jets}\xspace}
\newcommand{\al}{\ensuremath{\alpha}}
\newcommand{\alt}{\ensuremath{\alpha_{\text{T}}}\xspace}
\newcommand{\etaabs}{\ensuremath{|\eta|}}
\newcommand{\mjj}{\ensuremath{M_{\text{inv}}^{j1,j2}}}
\newcommand{\chiznew}{\ensuremath{\chi^{0}}\xspace}
\newcommand{\chipnew}{\ensuremath{\chi^{+}}\xspace}
\newcommand{\sQuanew}{\ensuremath{\tilde{\rm q}}\xspace}
\newcommand{\sGlunew}{\ensuremath{\tilde{\rm g}}\xspace}
\newcommand{\ttNew}{\ensuremath{\rm{t}\bar{\rm{t}}}\xspace}
\newcommand{\tev}{\TeV}
\newcommand{\combIso}{Iso_{\textrm{comb.}}}
\renewcommand{\arraystretch}{1.2}
\newcommand{\bigNum}[2]{#1 \, \times \, 10 \, ^{#2}}

\newcommand{\raT}{\ensuremath{R_{\alt}}}
\newcommand{\RaT}{\ensuremath{R_{\alt}}\xspace}

\def\eslash{{\hbox{$E$\kern-0.6em\lower-.05ex\hbox{/}\kern0.10em}}}  
\def\vecmet{\mbox{$\vec{\eslash}_T$}} 
\def\vecet{\mbox{$\vec{E}_\text{T}$}} 
\def\MET{\mbox{$\eslash_\text{T}$}\xspace}
\def\met{\mbox{$\eslash_\text{T}$}\xspace}
\def\mex{\mbox{$\eslash_\text{x}$}} 
\def\mey{\mbox{$\eslash_\text{y}$}} 
\def\mepar{\mbox{$\eslash_\parallel$}}
\def\meperp{\mbox{$\eslash_\perp$}}
\def\Zmm{Z \rightarrow \mu\mu}
\def\metvec{\mbox{$\vec{\met}$}\xspace}
\def\metvecrec{\mbox{$\vec{\met}^{\rm rec}$}\xspace}
\def\metvecgen{\mbox{$\vec{\met}^{\rm gen}$}\xspace}
\def\metgen{\mbox{$\met^{\rm gen}$}\xspace}
\def\metparl{\mbox{$\mepar^{\rm rec}$}\xspace}
\def\metperp{\mbox{$\meperp^{\rm rec}$}\xspace}
\def\deltamet{\mbox{$\Delta\met$}\xspace}
\def\pthat{\mbox{$\hat{p}_T$}\xspace}
\def\hslash{{\hbox{$H$\kern-0.8em\lower-.05ex\hbox{/}\kern0.10em}}}
\def\MHT{\mbox{$\hslash_\text{T}$}\xspace}
\def\mht{\mbox{$\hslash_\text{T}$}\xspace}
\def\sumet{\mbox{$\sum \rm{E}_\text{T}$}\xspace}
\def\scalht{\mbox{$H_\text{T}$}\xspace}
\def\etmiss{\mbox{$\eslash_\text{T}$}\xspace}
\def\htmiss{\mbox{$\hslash_\text{T}$}\xspace}
\def\mtt{\mbox{$\rm{M}_\text{T2}$}\xspace}
\def\rmec{\mbox{$R_{\mht/\met}$}\xspace}
\def\bdphi{\mbox{$\Delta\phi^{*}$}\xspace}
\def\bigeslash{{\hbox{$E$\kern-0.38em\lower-.05ex\hbox{/}\kern0.10em}}}  
\def\bigmet{\mbox{$\bigeslash_T$}}
\def\bighslash{{\hbox{$H$\kern-0.6em\lower-.05ex\hbox{/}\kern0.10em}}}  
\def\bigmht{\mbox{$\bighslash_T$}} 
\def\incl{\includegraphics[width=0.49\linewidth]}
\def\inclrot{\includegraphics[angle=90,width=0.47\linewidth]}
\def\INCL{\includegraphics[angle=90,width=0.45\linewidth]}
\def\Incl{\includegraphics[angle=90,width=0.60\linewidth]}

\def\Nch{\ensuremath{\left<N_{\text{ch}}\right>}\xspace}
\def\deltaR{\ensuremath{\left<\delta R^2\right>}\xspace}

\newcommand{\ppbar}{\ensuremath{\cmsSymbolFace{p}\overline{\cmsSymbolFace{p}}}\xspace} 
\cmsNoteHeader{QCD-10-029} 
\title{Shape, transverse size, and charged-hadron multiplicity of jets in pp collisions at \texorpdfstring{$\sqrt{s} = 7\TeV$}{sqrt(s) = 7 TeV}}

\date{\today}

\abstract{
Measurements of jet characteristics from inclusive jet production
in proton-proton collisions at a centre-of-mass energy of $7\TeV$
are presented.
The data sample was
collected with the CMS detector at the LHC during 2010
and corresponds to an integrated luminosity of $36\pbinv$.
The mean charged-hadron multiplicity,
the differential and integral jet shape distributions, and
two independent moments of the shape distributions are measured
as functions of the jet transverse momentum
for jets reconstructed with the anti-$k_{\mathrm{T}}$ algorithm.
The measured observables are corrected to the particle level and compared
with predictions from various QCD Monte Carlo generators.
}

\hypersetup{%
pdfauthor={CMS Collaboration},%
pdftitle={Shape, transverse size, and charged hadron multiplicity of jets in pp collisions at 7 TeV},%
pdfsubject={CMS},%
pdfkeywords={CMS, physics, QCD, jets}}

\maketitle

\section{Introduction}
The jet transverse momentum profile (shape)~\cite{Ellis:1992qq,Seymour:1997kj}, 
transverse size,
and charged-hadron multiplicity in jets are sensitive to multiple parton 
emissions from the primary outgoing parton and provide a powerful test of the 
parton showering approximation of quantum chromodynamics (QCD), the theory of 
strong interactions. Recently, there have been many methods proposed to search for
heavy particles by studying the substructure of jets formed by their decay products, as these particles can
be highly boosted and thus their decay products are well collimated
~\cite{Seymour:1993mx,Butterworth:2008iy,Kaplan:2008ie,Ellis:2009su}. 
Jets arising from the fragmentation of a single parton, hereafter referred to as QCD jets, 
contribute to backgrounds in searches for such boosted-object jets.
A good understanding of the QCD jet structure is very important for these searches to be successful.  
The structures of gluon-initiated and quark-initiated
jets are different due to their different fragmentation properties.
QCD predicts gluon-initiated jets to have a higher average
particle multiplicity and a broader distribution of particle transverse
momentum with respect to the jet direction compared to quark-initiated jets. 
Jet structure measurements test these predictions and can be used to develop techniques   
to discriminate between gluon and quark jets. Such discrimination techniques can enhance both
standard model measurements and the ability to search for physics beyond the standard model.

Historically the jet shape has been used to test perturbative QCD (pQCD) calculations up to the third power in the coupling 
constant $\alpha_s$~\cite{Abe:1992wv,Abachi:1995zw}. These leading-order calculations, 
with only one additional parton in a jet, showed reasonable agreement with the 
observed jet shapes. While confirming the validity of pQCD calculations, jet 
shape studies also indicated that jet clustering, underlying event contributions,
and hadronization effects must be considered. 
Currently, these
effects are modelled within the framework of Monte Carlo (MC) event generators, which 
use QCD parton shower models, in conjunction with hadronization and underlying
event models, to generate final-state particles. These MC event generators are used extensively
to model the signal and background events for a variety of standard model studies and searches
for new physics at hadron colliders. Jet shapes are used to tune phenomenological parameters
in the event generators.
Jet shapes have been measured previously in 
$\cmsSymbolFace{p}\overline{\cmsSymbolFace{p}}$
collisions at the Tevatron~\cite{Abe:1992wv,Abachi:1995zw,Acosta:2005ix,CDFBJetPaper}
and $\cmsSymbolFace{ep}$ collisions at HERA~\cite{Aid:1995ma,Adloff:1997gq,Breitweg:1997gg,Adloff:1998ni,Breitweg:1998gf}.

We present measurements of the charged-hadron multiplicity, shape, 
and transverse size for jets with transverse momentum up to $1$\TeV and rapidity up to 3 using 36\pbinv of 
$\cmsSymbolFace{pp}$ collisions 
at a centre-of-mass energy of $7$\TeV collected by the CMS experiment at the Large Hadron Collider (LHC). 
A similar measurement has been performed by the ATLAS Collaboration~\cite{ATLAS_jetshape}. 

This paper is organised as follows.  
Section~2 contains a brief description of the CMS detector.  
In Section~3 we present the event selection and reconstruction.
The jet observables are defined in Section~4 and the results are given in Section~5.  
The conclusions are summarized in Section~6.

\section{The CMS detector}
CMS uses a right-handed coordinate system in which the $z$ axis 
points in the anticlockwise beam direction,
the $x$ axis points towards the centre of the LHC ring, 
and the $y$ axis points up, perpendicular to the plane of the LHC ring. 
The azimuthal angle $\phi$ is measured in radians 
with respect to the $x$ axis, and the polar angle $\theta$ is measured with respect to the $z$ axis. 
A particle with energy $E$ and momentum $\vec{p}$ is characterized by
transverse momentum $\pt = |\vec{p}|\,\sin{\theta}$,
rapidity $y= \frac{1}{2} \ln\left[ (E+p_{z})/(E-p_{z}) \right]$, and
pseudorapidity $\eta = -\ln\left[ \tan(\theta/2) \right]$.

The CMS superconducting solenoid, $12.5$ m long
with an internal diameter of $6$ m, provides a uniform magnetic field of
$3.8$ T.
The inner tracking system is composed of a pixel detector with
three barrel layers at radii between $4.4$ and $10.2$ cm
and a silicon strip tracker with 10 barrel detection layers extending
outwards to a radius of $1.1$ m. This system is complemented by two
endcaps, extending the acceptance up to $|\eta|=2.5$. 
The momentum resolution for reconstructed tracks 
in the central region is about 1\% at $\pt$ = 100\GeVc. 

The calorimeters inside the magnet coil consist of a
lead tungstate crystal electromagnetic calorimeter (ECAL)  and a
brass-scintillator hadron calorimeter (HCAL) with coverage up
to $|\eta|=3$.
The quartz/steel forward hadron calorimeters extend the calorimetry coverage up to $|\eta|=5$.
Muons are measured in gas-ionization detectors embedded in the steel
return yoke of the magnet.
The calorimeter cells are
grouped in projective towers of granularity
$\Delta\eta\times\Delta\phi = 0.087 \times 0.087 $
for the central rapidities considered in this paper.
The ECAL was initially calibrated using test beam electrons and then, {\it{in situ}},
with photons from $\pi^0$ and $\eta$ meson decays and electrons
from \Z boson decays~\cite{CMS-DP-2011-008}.
The energy scale in data agrees with that in the simulation to better than 1\% in the barrel region ($|\eta|<1.5$) 
and better than 3\% in the endcap region ($1.3<|\eta|<3.0$)~\cite{pas_ecal}.
Hadron calorimeter cells in the $|\eta|<3$ region are calibrated primarily with test-beam data and
radioactive sources~\cite{hcal_jinst,HcalTestBeam}. 
A detailed description of the CMS detector may be found in~\cite{CMS_Detector}.

\section{Event selection and reconstruction}
The data were recorded using a set of inclusive single-jet high-level triggers~\cite{HLT}
requiring at least one jet in the event to have an online jet \pt
of at least 15, 30, 50, 70, 100, or 140\GeVc. 
These jets are reconstructed only from energy deposits in the calorimeters
using an iterative cone algorithm.
In addition, a minimum-bias trigger, 
defined as a signal from at least one of two beam scintillator counters in coincidence
with a signal from one of two beam pickup timing devices,
was used to collect low \pt jets.
These datasets are combined to measure the jet characteristics in bins 
spanning the range 20\GeVc $<\pt<$ 1\TeVc{}, so that 
the trigger contributing to each bin is fully efficient. 
Only a fraction of  events satisfying the lower threshold jet triggers were recorded
because of limited data acquisition system bandwidth. 
Thus the effective integrated luminosity for jets with $\pt<140$\GeVc is less than 36 \pbinv.

Jets are reconstructed offline using the anti-\kt jet clustering algorithm~\cite{antikt,Cacciari:fastjet1,Cacciari:fastjet2}.
This algorithm is similar to the well-known \kt algorithm, except that it uses $1/\pt$ instead 
of \pt as the weighting factor for the scaled distance. The algorithm is collinear- and infrared-safe, 
and it produces circular jets  in $y$-$\phi$ space except when jets overlap. 
Two different types of inputs are used with this algorithm.
In the first method, 
individually calibrated particle candidates are used as inputs to the jet clustering algorithm.
These particle candidates, photons, electrons, muons, charged hadrons, and neutral hadrons,
are reconstructed using the CMS particle flow (PF) algorithm~\cite{pas_ParticleFlow}.
This algorithm combines the information from all the subdetectors including the silicon tracking system,
the electromagnetic calorimeter, 
the hadron calorimeter, 
and the muon system in order to reconstruct and identify individual particles in an event.
The charged-particle information is primarily derived from the tracking system, and the
photons are reconstructed using information from the electromagnetic calorimeter. 
The neutral hadrons, e.g.\ neutrons and \PKzL\ mesons,
carry on average about 15\% of the jet momentum, 
and are reconstructed using information from the hadron calorimeter. 
Jets reconstructed from these inputs are referred to as PF jets.

In the second method, called the jet-plus-track (JPT) algorithm~\cite{JPTJets},
the energy deposits in the electromagnetic and hadron calorimeter cells, which are combined into calorimeter towers,
are used as inputs to the clustering algorithm to form calorimeter jets.
Tracks originating from the interaction vertex~\cite{pas_tracking} are associated with these calorimeter 
jets based on the separation in $\eta$-$\phi$ space between the jet direction and track direction at the
interaction vertex. In the case of partially overlapping jets, tracks are assigned to the jet with the minimum 
\pt-weighted distance 
between each track and the jet axis. 
These tracks are categorized as muon, charged pion, and electron candidates, and the
jet momentum is corrected by substituting their expected particle energy deposition in the calorimeter with their momentum. 
These track-corrected jets are referred to as JPT jets.

The \pt of both types of jets are corrected to the particle-level jet \pt{}~\cite{JETJINST}.
In both cases, 
the ratio of the reconstructed jet \pt
to the particle jet \pt is close to unity, and only small additional
corrections to the jet energy scale, of the order of 5--10\%,  are needed.
These corrections are derived from GEANT4-based~\cite{Agostinelli2003250} 
CMS simulations, based on the \pt ratio of the particle jet formed
from all stable ($c\tau>1$ cm) particles to the reconstructed jet,
and also {\it{in situ}}
measurements using dijet and photon + jet events~\cite{JETJINST}.
The uncertainty on the absolute jet energy scale is studied using both data and MC events 
and is found to be less than 5\% for all values of jet \pt and $\eta$.
In order to remove jets coming from instrumental noise,
jet quality requirements are applied~\cite{JetMet}.

The JPT jets are reconstructed with the anti-\kt jet clustering algorithm and
distance parameter $D=0.5$~\cite{antikt}. 
The tracks associated with JPT jets are used to measure the charged-hadron
multiplicity and the transverse size of the jets in the jet \pt range 50\GeVc $<\pt<$ 1\TeVc.
The PF jets reconstructed with a distance parameter $D=0.7$
are used to measure the jet shapes in the jet \pt range 20\GeVc $<\pt<$ 1\TeVc.
Owing to the larger jet size, jet shape measurements evaluate
a larger fraction of the momentum from the originating parton and are relatively more sensitive to momentum deposited
by multiple-parton interactions (MPIs), thus providing important information to tune both the parton showering and 
MPI models in the event generators.
To minimize the contribution from additional $\cmsSymbolFace{pp}$ interactions in a triggered event (pileup), 
events with only one reconstructed primary vertex are selected for jet shape measurements, 
as the measurements use both charged and neutral particles. 
For charged-hadron multiplicity and jet transverse size studies, the events with multiple vertices are also considered as these studies use
only those tracks that are associated with the primary vertex.
The primary vertex is defined as the vertex with the highest sum of transverse momenta
of all reconstructed tracks pointing to it.

\section{Jet observables}
We have studied several observables to characterize the jet structure.
These observables are complementary and they can provide
a more comprehensive picture of the composition of jets. 
In order to compare the resulting measurements with
theoretical predictions, all the observables are corrected back to the particle
level by taking into account detector effects using MC simulations.

\subsection{Jet shapes} 
\begin{figure}[htb]
\begin{center}
\includegraphics*[width=2.5in, height=2.5in,viewport=200 350 500 650]{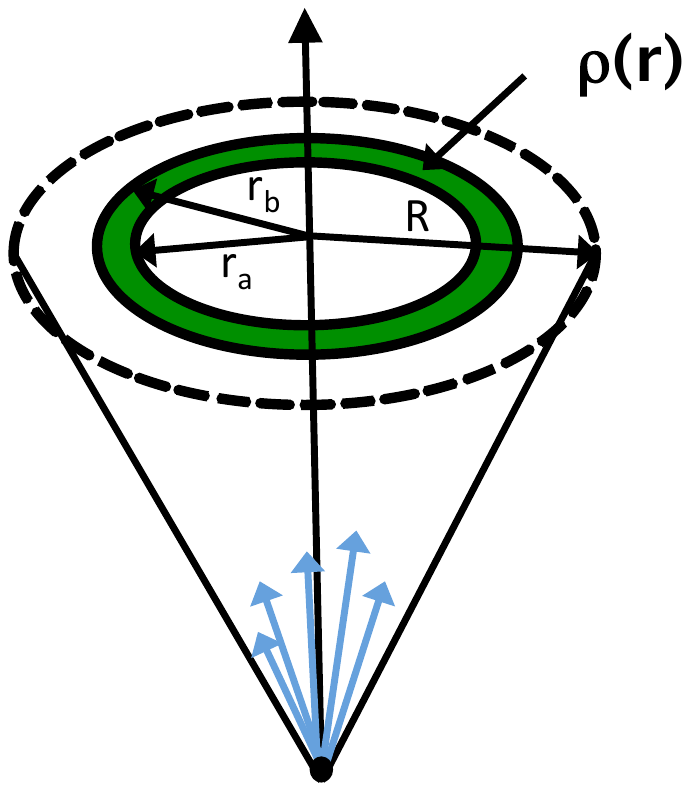}
\includegraphics*[width=2.5in,height=2.5in,viewport=200 350 500 650]{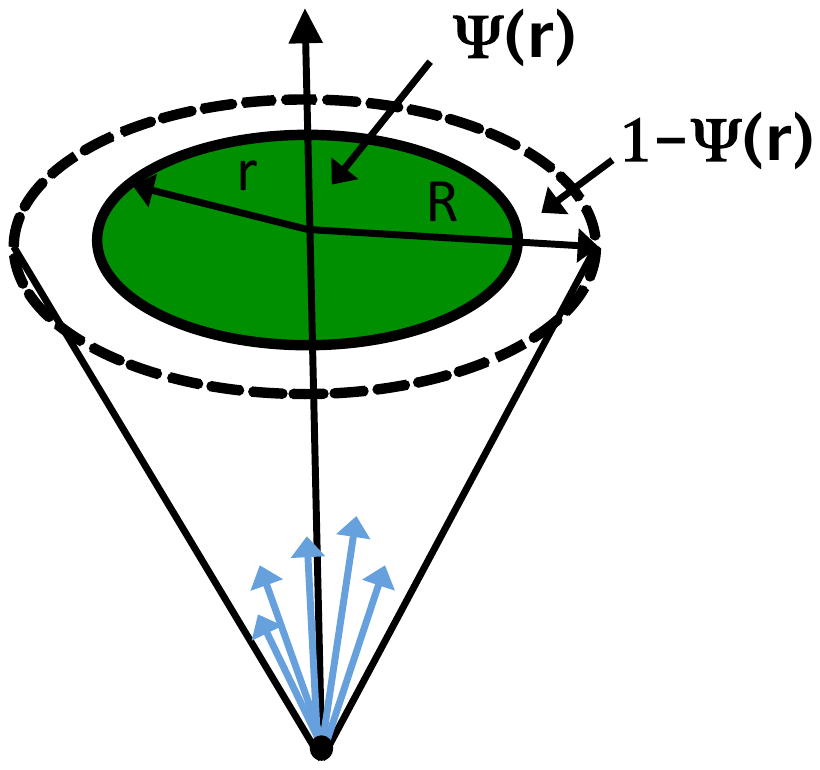}
  \caption{Pictorial definition of the differential (top) and integrated (bottom) jet shape quantities. 
    Analytical definitions of these quantities are given in the text.
  }
  \label{Jetshape_def}
\end{center}
\end{figure}
The differential jet shape  $\rho(r)$ is defined as the average fraction of
the transverse momentum contained inside an annulus
of inner radius $r_{\mathrm{a}}= r-\delta r/2$ and outer radius $r_{\mathrm{b}}=r+\delta r/2$ 
as illustrated in Fig.~\ref{Jetshape_def}:
$$
     \rho(r) = \frac{1}{\delta r} 
               \frac{\sum\limits_{r_{\mathrm{a}} < r_{{i}} <r_{\mathrm{b}}}\; p_{\mathrm{T},i}}
                    {\sum\limits_{r_{{i}} < R} \; p_{\mathrm{T},i}},
$$
where $\delta r = 0.1$.

The integrated jet shape $\Psi(r)$ is defined as the average fraction of the
transverse momentum of particles inside a cone of radius
$r$  around
the jet axis:
$$
     \Psi(r) =      \frac{\sum\limits_{r_{{i}} <r}\; p_{\mathrm{T},i}}
                    {\sum\limits_{r_{{i}} < R} \; p_{\mathrm{T},i}}.
$$
The sums run over the reconstructed particles, with the distance
$r_{{i}} =\sqrt{(y_{{i}}-y_{\text{jet}})^2 + (\phi_{{i}}-\phi_{\text{jet}})^2}$
relative to the jet axis described by $y_{\text{jet}}$ and $\phi_{\text{jet}}$,
and $R=0.7$. 

The observed detector-level jet shapes and true particle-level jet shapes differ because of
jet energy resolution effects, 
detector response to individual particles, 
smearing of the jet directions,
smearing of the individual particle directions, and 
inefficiency of particle reconstruction, especially at low \pt.
The data are unfolded to the particle level using bin-by-bin corrections
derived from the CMS simulation
based on the {\sc pythia} 6.4 ({\sc pythia6}) MC generator~\cite{Pythia} tuned to the CMS
data (tune Z2). The Z2 tune is identical to the Z1 tune described in~\cite{Field}, except that Z2 uses the CTEQ6L~\cite{Pumplin:2002vw}
parton distribution function (PDF), while Z1 uses CTEQ5L~\cite{CTEQ5} PDF. 
The correction factors are determined as functions of $r$ for each jet \pt and rapidity bin
and vary between 0 and 20\%.
Since the MC model affects the momentum and angular distributions and flavour composition of particles in a jet, and 
therefore the simulated detector response to the jet, the unfolding factors depend on the MC model.
In order to estimate the systematic uncertainty due to the fragmentation model,
the corrections are also derived using {\sc pythia8}~\cite{Pythia8}, {\sc pythia6} tune D6T~\cite{Pythia},
and {\sc herwig++}~\cite{Bahr:2008pv}.
The largest difference of these three sets of correction factors from those of {\sc pythia6} tune Z2
is assigned as the uncertainty on the correction. 
This uncertainty is typically 2--3\% in the region where the bulk of the jet energy is deposited
and increases to as high as 15\% at large radii where the momentum of particles is very small. 
For very high \pt jets where the fraction of jet momentum deposited at large radii is extremely small, the uncertainty
is less than 1\% at $r=0.1$ and reaches 25\% at high radii.

The impact of the calibration uncertainties for particles used to measure the jet shapes
is studied separately for charged hadrons, neutral hadrons, and photons.
The calibration of each type of particle is varied within its measurement uncertainty, 
depending on its \pt and $\eta$. 
The resulting change in the jet shape distributions is negligible
as expected since the effect is largely cancelled out in the jet shapes, which are defined as \pt sum ratios. 

The jet energy scale uncertainty has a larger impact on the jet shape
measurements because it affects the migration of jets 
between different jet \pt bins.
The jet energy scale uncertainty is estimated
to be less than 5\% for all jet \pt and $\eta$ bins~\cite{JETJINST} and results in a
maximum uncertainty of 2--3\% in both the differential and integrated jet shape distributions.

\subsection{The charged-hadron multiplicity and the transverse size of jets}
In addition to the study of $\rho(r)$ and $\Psi(r)$, 
we have measured characteristics of the charged components of jets,
namely, the mean charged-hadron multiplicity per jet, \Nch, and the second moments of the transverse
jet size, defined by
    $$
      \left< \delta \eta^2 \right> 
       = \frac{\sum\limits_{{i} \in \text{jet}} 
                 \left(\eta_{{i}} -\eta_{\mathrm{C}} \right)^2 \, \cdot p_{\mathrm{T},i}}
              {\sum\limits_{{i} \in \text{jet}} p_{\mathrm{T},i}}\;,
     \;\;\;\;\;\;\;\;\;\;\
      \left< \delta \phi^2 \right> 
       = \frac{\sum\limits_{{i} \in \text{jet}}
                 \left(\phi_{{i}}-\phi_{\mathrm{C}} \right)^2 \, \cdot p_{\mathrm{T},i}}
              {\sum\limits_{{i} \in \text{jet}} p_{\mathrm{T},i}}\;,
    $$
where
    $$
      \eta_C 
       = \frac{\sum\limits_{{i} \in \text{jet}} \eta_{{i}} \cdot p_{\mathrm{T},i}}
              {\sum\limits_{{i} \in \text{jet}} p_{\mathrm{T},i}} \;,
     \;\;\;\;\;\;\;\;\;\;\
     \;\;\;\;\;\;\;\;\;\;\
      \phi_C 
       = \frac{\sum\limits_{{i} \in \text{jet}} \phi_{{i}} \cdot p_{\mathrm{T},i}}
              {\sum\limits_{{i} \in \text{jet}} p_{\mathrm{T},i}} \;,
    $$

and $p_{\mathrm{T},i}$, $\eta_{{i}}$, and $\phi_{{i}}$ are the transverse momentum, 
pseudorapidity, and azimuthal direction of a
particle $i$ in the jet. 
These moments are combined to obtain the second moment of the jet transverse width:
    $$ \deltaR  = 
     \left<\delta \eta ^{2}\right> + \left<\delta \phi ^{2}\right>\;.
    $$ 

We measure \Nch and \deltaR using tracks with $\pt>0.5$\GeVc associated with JPT jets.
The tracks identified as electrons or muons are explicitly removed. As the tracks are required to be attached to
the primary vertex, the tracks resulting from photon conversions are not used either.

The particle-level \Nch and \deltaR values, defined to correspond to all stable charged hadrons with $\pt>0.5$\GeVc{}, are obtained 
by separately correcting the measured observables for the tracking inefficiency and the jet energy resolution.
The corrections to the track detection efficiency are applied in two steps:
first, corrections for the tracker acceptance
and for losses due to interactions in the detector material are determined for isolated charged pions as functions of \pt and $\eta$
using CMS simulation and applied as a weight assigned to each track~\cite{TRK-10-002}.
Next, residual corrections for both the tracking inefficiency and misidentified tracks 
inside the dense high-\pt jet environment are calculated for \Nch and 
\deltaR as functions of jet \pt in two jet rapidity ranges: $|y|<1$ and $1<|y|<2$. 
These corrections are derived from MC by comparing the 
detector-level and particle-level \Nch and \deltaR for each jet \pt bin. The correction factors for \Nch increase 
from about 2\% for jets with $\pt=40\GeVc$  to 5\% for jets with $\pt=200\GeVc$.
The corrections increase to 20\% for a jet with \pt of 800\GeVc.
For \deltaR, the corrections increase from 3 to 8\% 
as the jet \pt goes from 40\GeVc to 200\GeVc, and rise to 20\% for 800\GeVc \pt jets. The uncertainty on \Nch due to
these residual corrections is 1\%, while the uncertainty on \deltaR is 2--5\%.  

The jet energy resolution corrections are extracted bin by bin from  
the CMS simulation based on the {\sc pythia6} tune Z2 MC samples. 
The uncertainty on \Nch (\deltaR) due to jet energy resolution is 1--2\% (2--5\%).
A cross-check of the correction procedure is performed using the Tikhonov regularization method with a quasi-optimal solution~\cite{Bogolyubsky:1995vr,tich2}.
The results obtained with these two methods are consistent to within 2\%.

\section{Results}
In this section, data results are compared to MC simulations using the
{\sc pythia6}~\cite{Pythia}, {\sc pythia8}~\cite{Pythia8}, 
and {\sc herwig++}~\cite{Bahr:2008pv} event generators.
Three different tunes of the {\sc pythia} 6.4 generator are considered:
tune D6T, tune Z2, and the Perugia2010 tune~\cite{PerugiaTunes}.
Tune D6T uses virtuality-ordered parton showers, while tune Z2 and the Perugia2010 tune
use $\pt$-ordered parton showers.
Generator input parameters controlling the underlying event, radiation, and hadronization  
are tuned in order to provide a better description of collider data. 
Tune D6T was developed using previous hadron and lepton collider data, while tune Z2 also uses 
CMS soft \pt data~\cite{UE_paper}. 
The Perugia2010 tune~\cite{PerugiaTunes} was tuned using LEP and Tevatron data, 
notably the CDF jet shape results~\cite{Acosta:2005ix}. 
Tunes D6T and Z2 are simulated with {\sc pythia} 6.4.22, and the Perugia2010 tune 
is simulated with {\sc pythia} 6.4.24.
The CTEQ6L1~\cite{Pumplin:2002vw} parton distribution function (PDF) of the proton 
is used with tunes D6T and Z2, while
the CTEQ5L~\cite{CTEQ5} PDF is used with the Perugia2010 tune.
The {\sc pythia} 8.145 generator, tune 2C, uses an improved diffraction model, and the 
{\sc herwig++} 2.4.2 generator uses angular-ordered parton showers and a cluster-based 
fragmentation model.
For {\sc herwig++} 2.4.2, the default underlying event tune is used together with the MRST2001~\cite{Martin:2001es} PDF.

\begin{figure}[p]
\begin{center}
\includegraphics[width=0.35\hsize,clip=false,viewport=0 0 550 650]{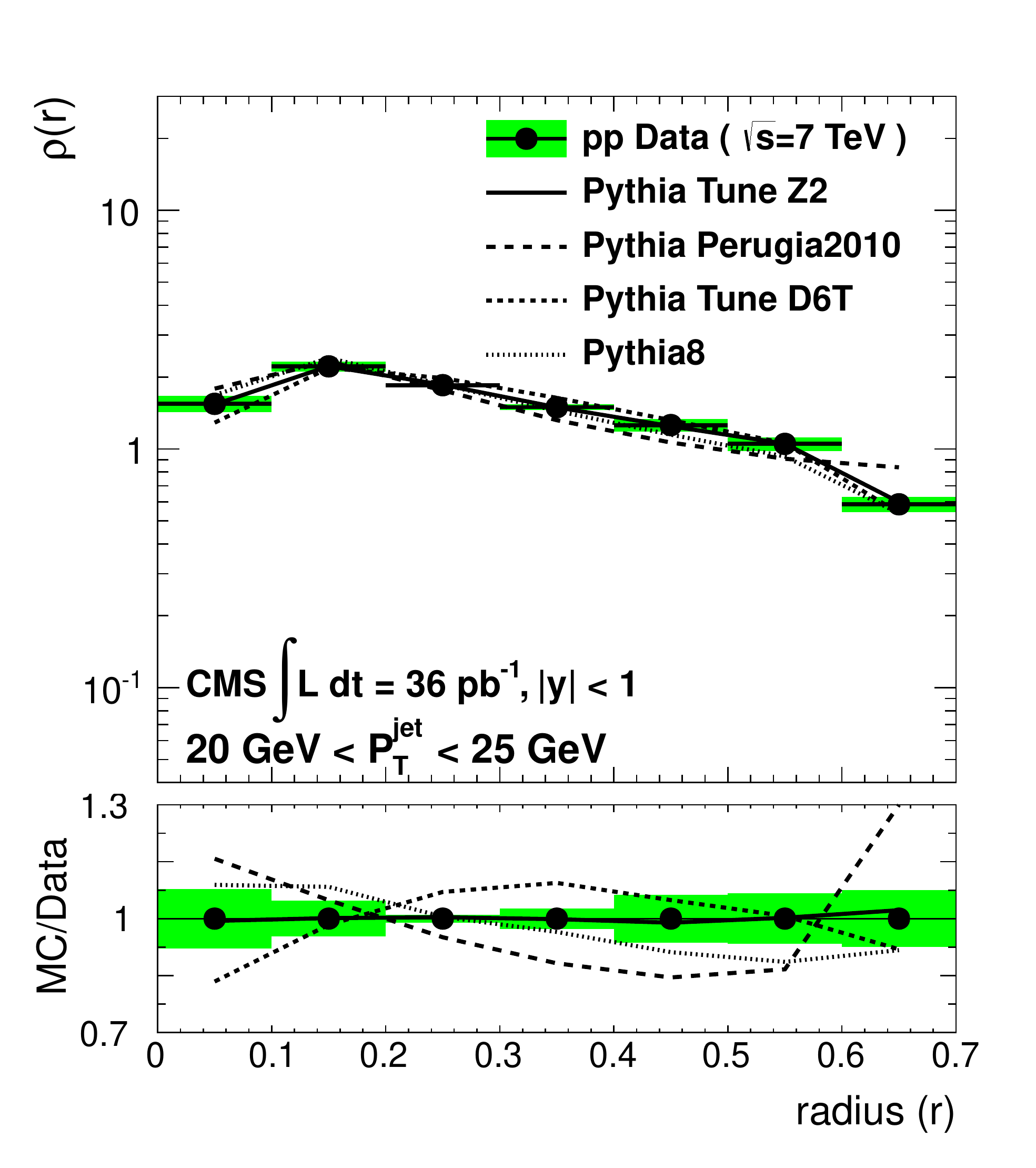}
\includegraphics[width=0.35\hsize,clip=false,viewport=0 0 550 650]{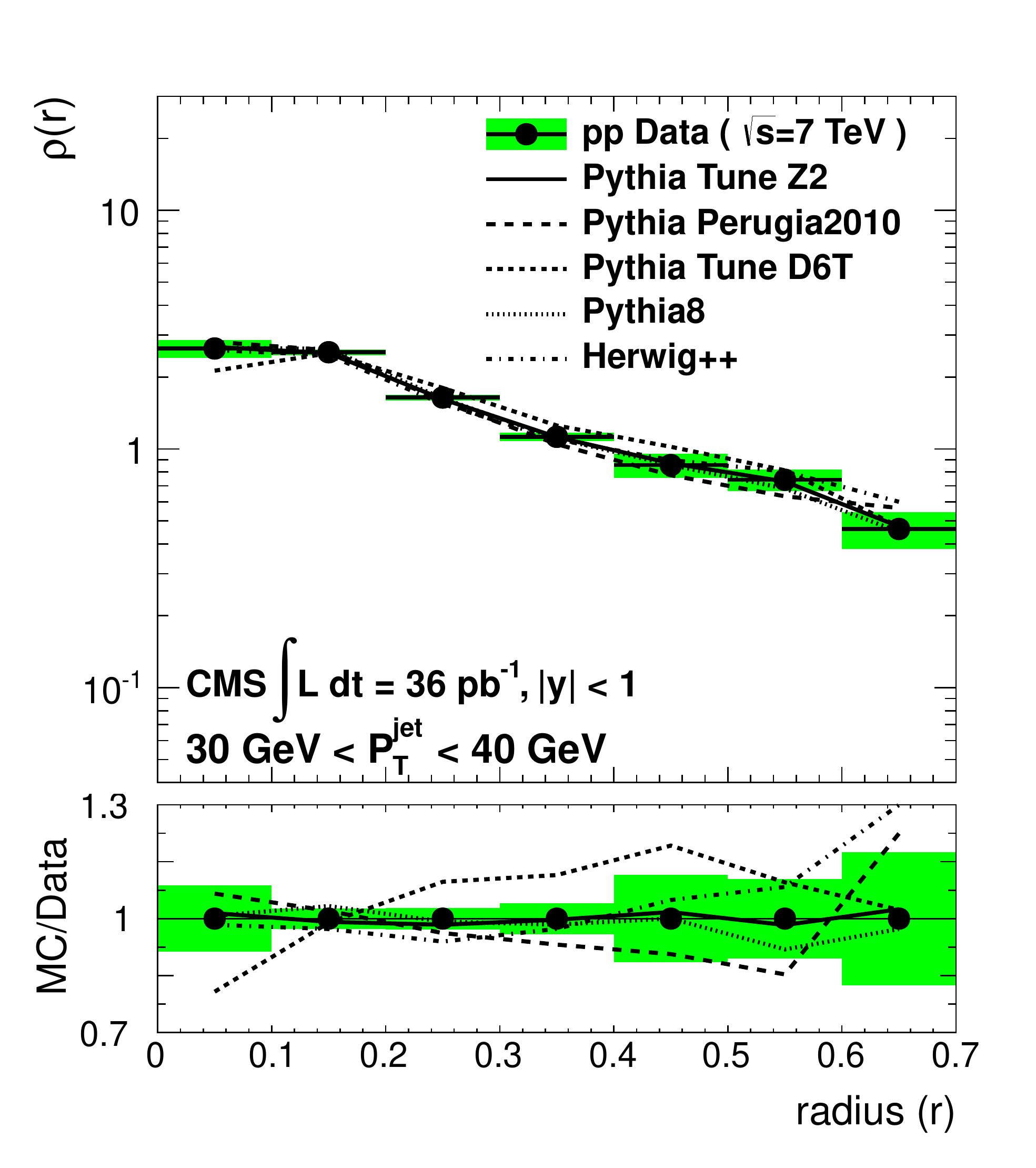}
\includegraphics[width=0.35\hsize,clip=false,viewport=0 0 550 650]{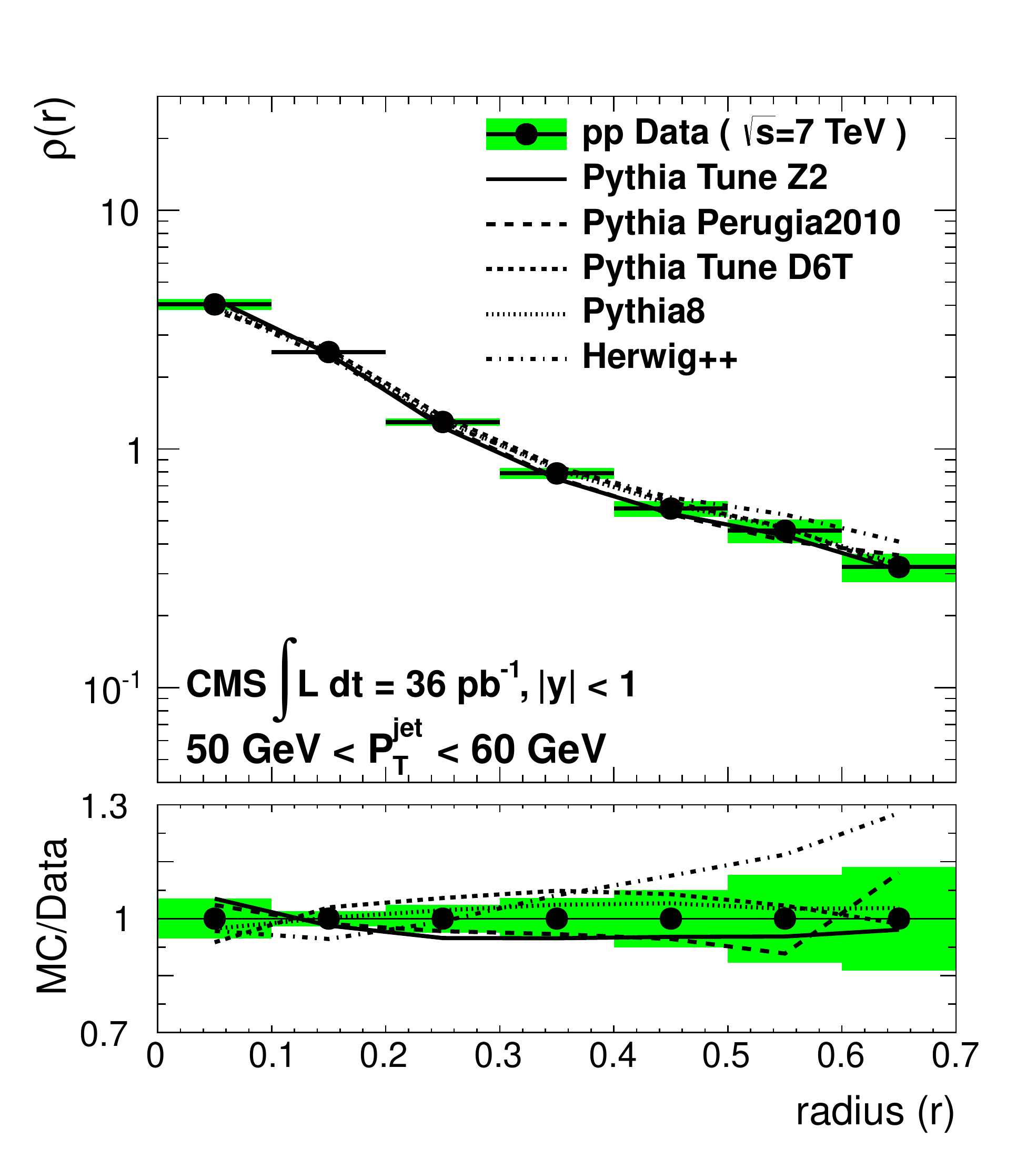}
\includegraphics[width=0.35\hsize,clip=false,viewport=0 0 550 650]{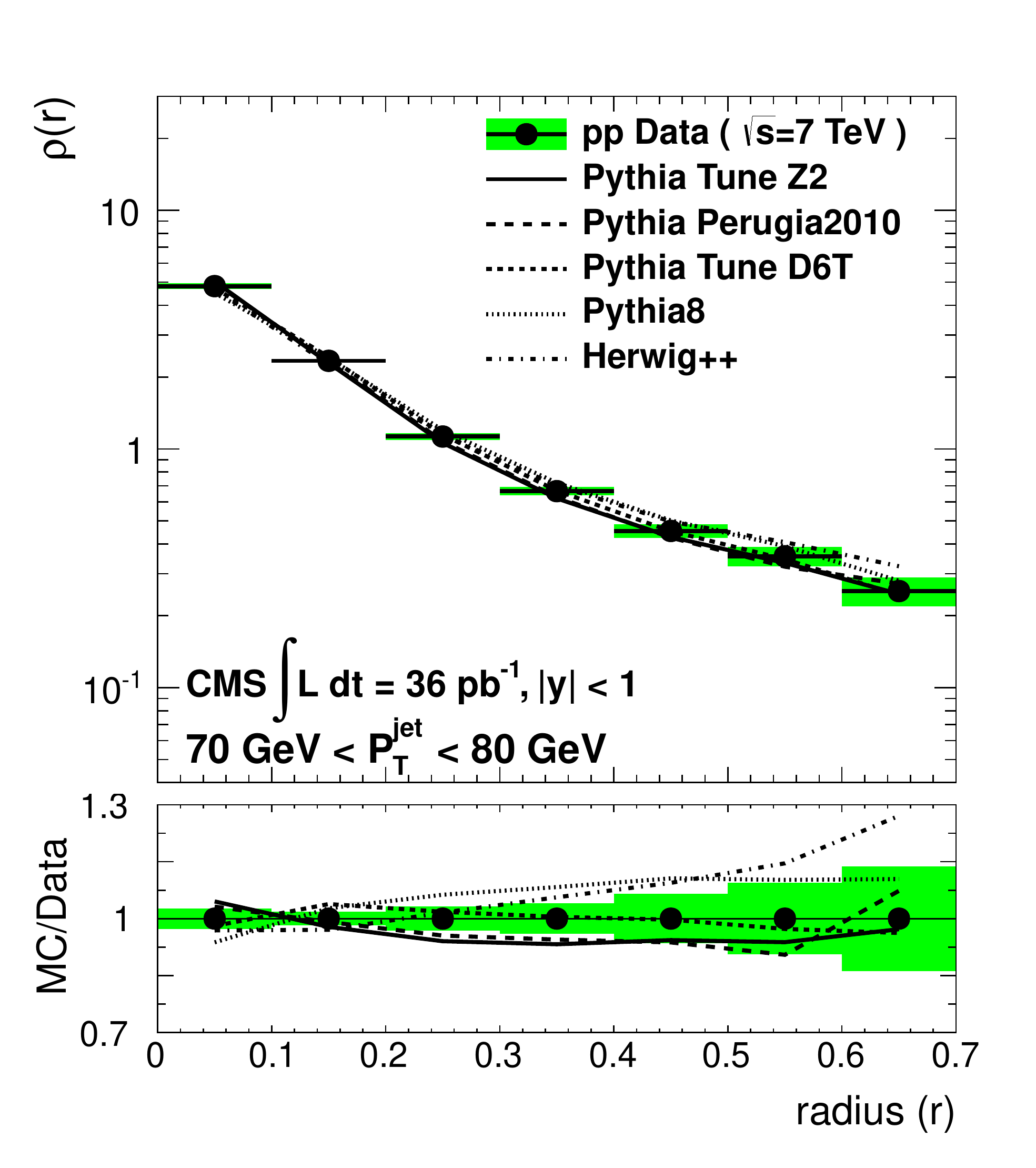}
\includegraphics[width=0.35\hsize,clip=false,viewport=0 0 550 650]{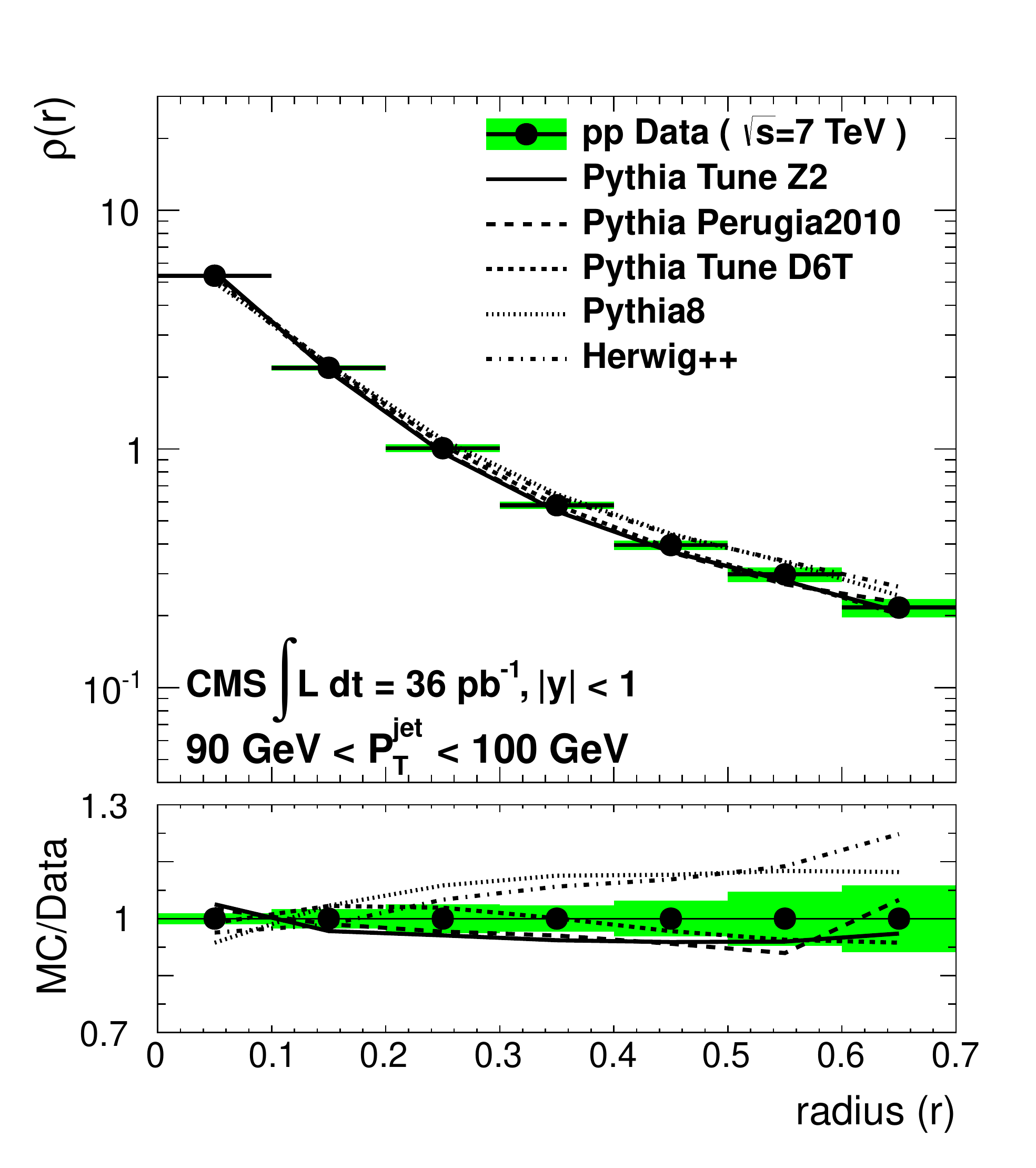}
\includegraphics[width=0.35\hsize,clip=false,viewport=0 0 550 650]{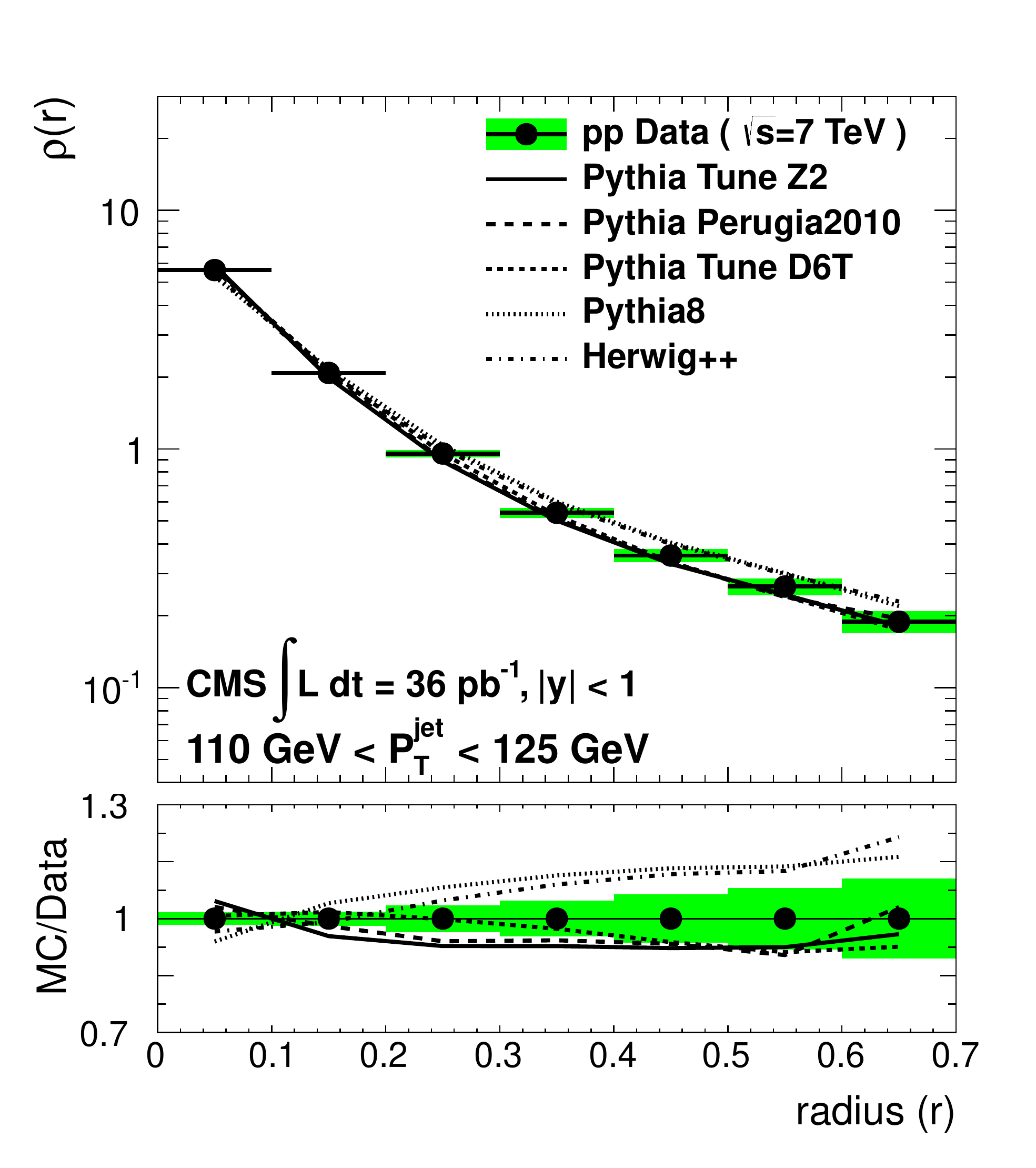}
\caption{Differential jet shape 
as a function of the distance from the jet axis for central jets ($|y|<1$) with jet 
transverse momentum ranging from 20 to 125\GeVc for representative jet $\pt$ bins. 
The data are compared to particle-level
{\sc herwig++}, {\sc pythia8}, and {\sc pythia6} predictions with various tunes. 
Statistical uncertainties are shown as error bars on the data points and
the shaded region represents the total systematic uncertainty of the measurement.
Data points are placed at the bin centre; the horizontal bars show the
size of the bin. 
The ratio of each MC prediction to the data is also shown in the lower part of each plot. 
}
\label{fig-Diff1}
\end{center}
\end{figure}

\begin{figure}[p]
\begin{center}
\includegraphics[width=0.35\hsize,clip=false,viewport=0 0 550 650]{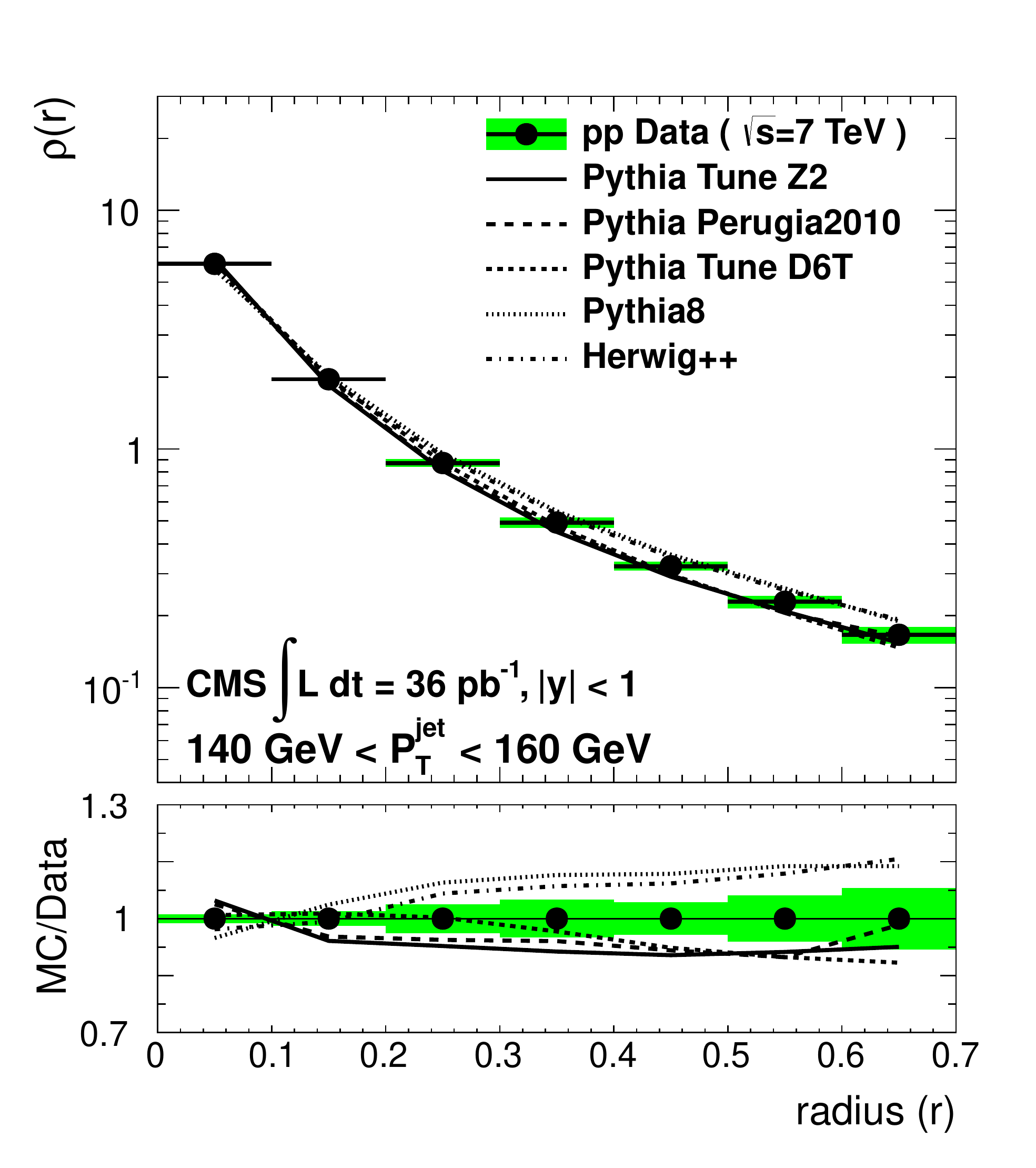}
\includegraphics[width=0.35\hsize,clip=false,viewport=0 0 550 650]{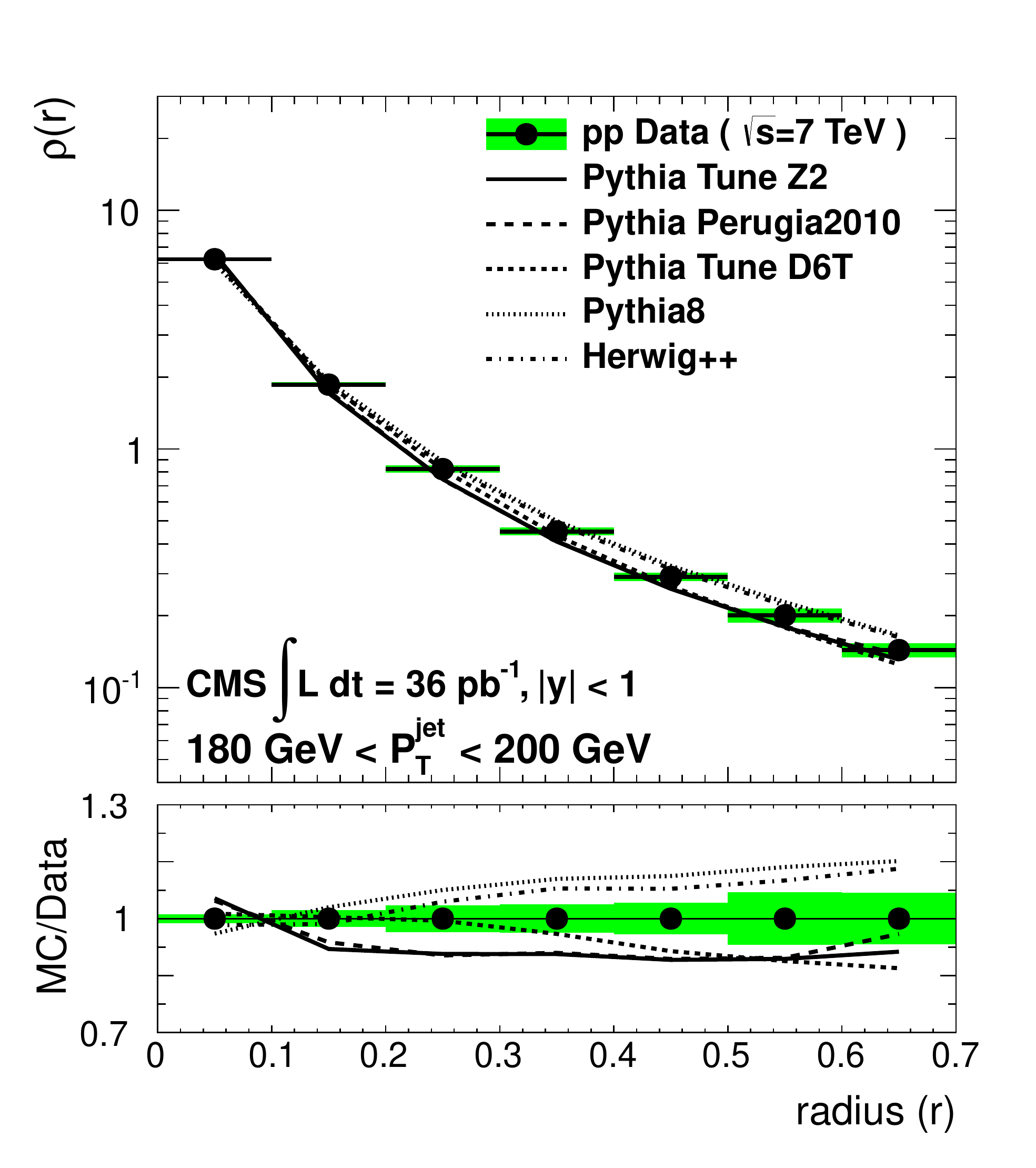}
\includegraphics[width=0.35\hsize,clip=false,viewport=0 0 550 650]{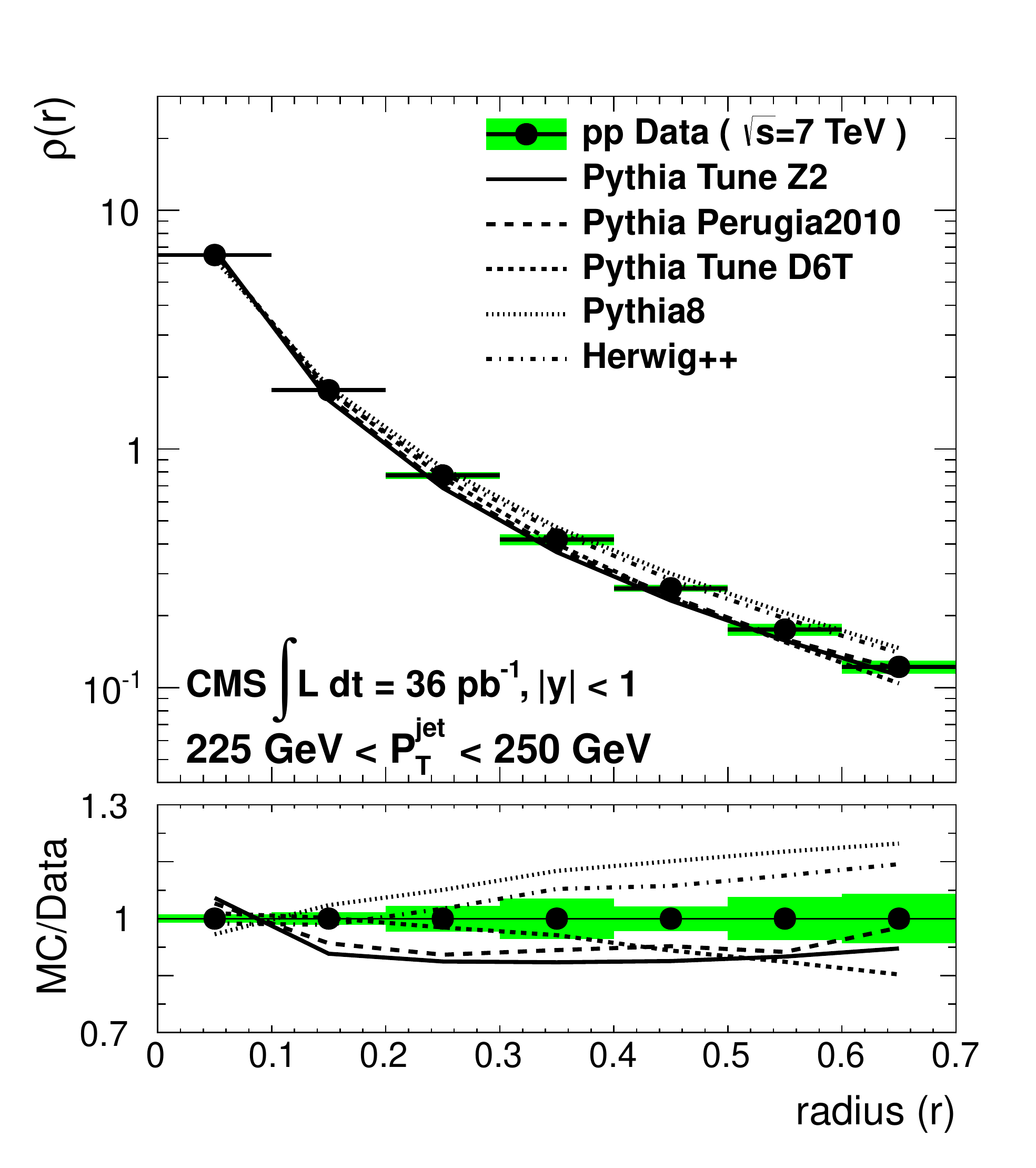}
\includegraphics[width=0.35\hsize,clip=false,viewport=0 0 550 650]{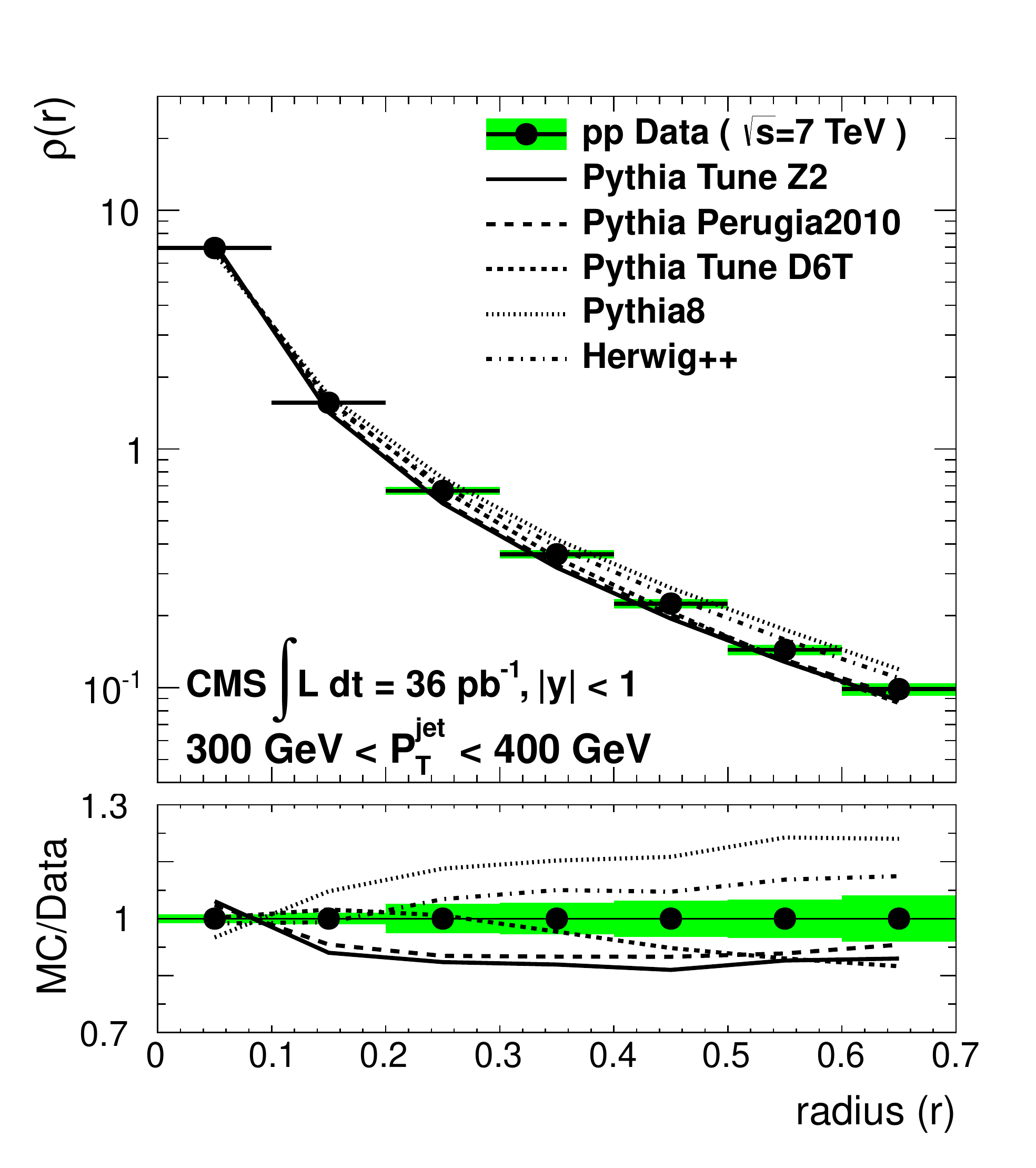}
\includegraphics[width=0.35\hsize,clip=false,viewport=0 0 550 650]{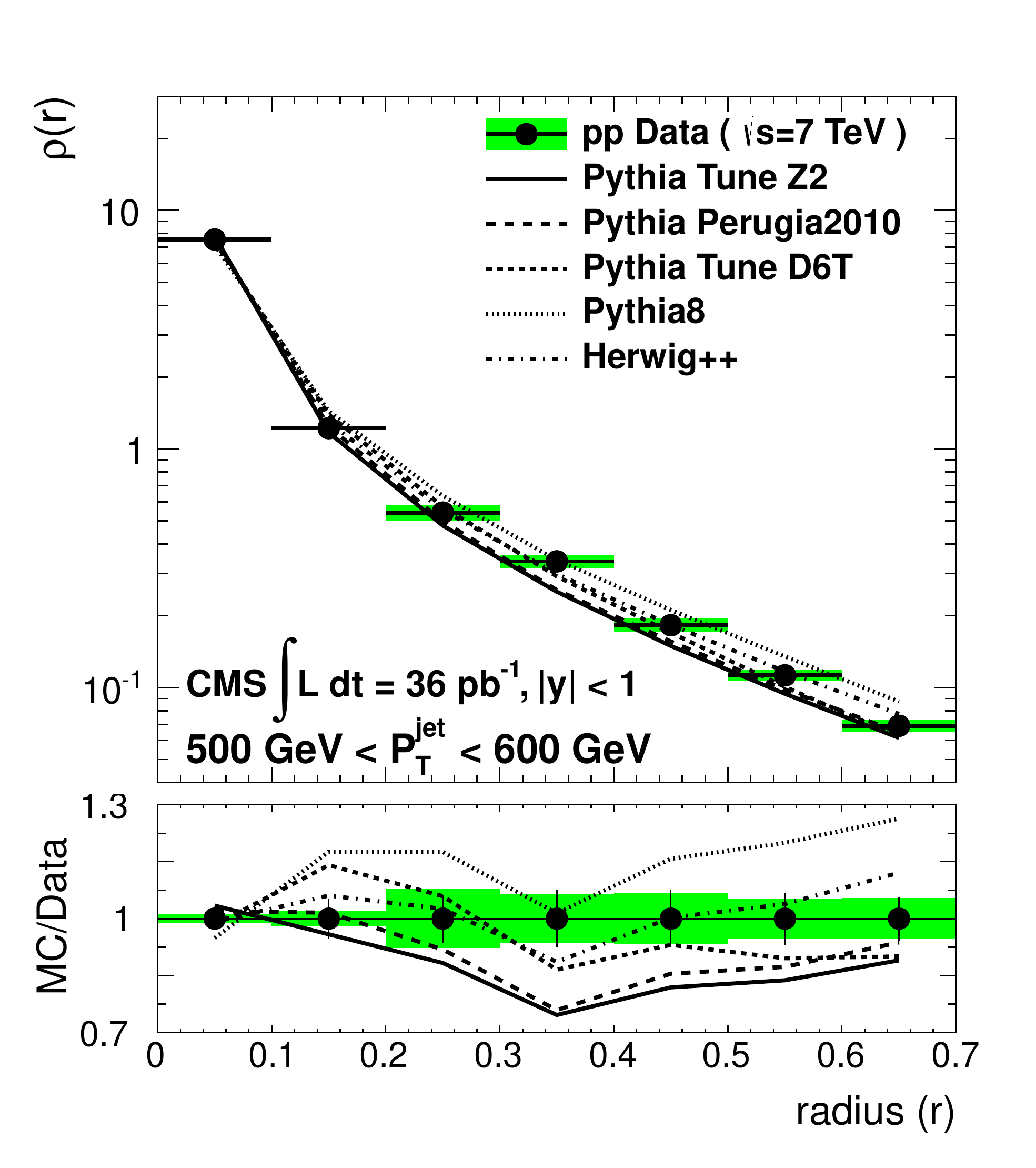}
\includegraphics[width=0.35\hsize,clip=false,viewport=0 0 550 650]{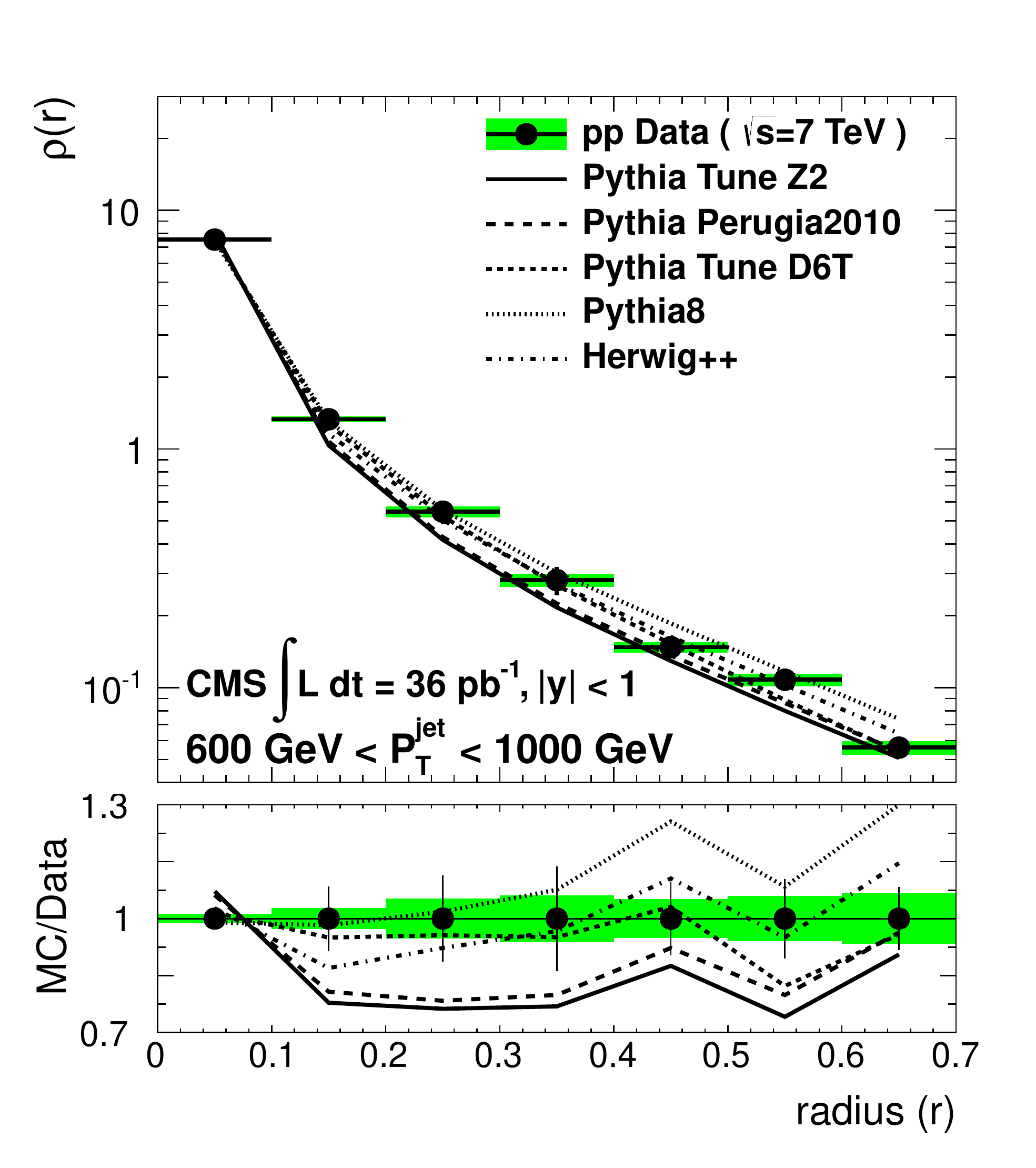}
\caption{Differential jet shape 
as a function of the distance from the jet axis for central jets ($|y|<1$) with jet
transverse momentum ranging from 140 to 1000\GeVc for representative jet $\pt$ bins. 
The data are compared to particle-level 
{\sc herwig++}, {\sc pythia8}, and {\sc pythia6} predictions with various tunes. 
Statistical uncertainties are shown as error bars on the data points and
the shaded region represents the total systematic uncertainty of the measurement.
Data points are placed at the bin centre; the horizontal bars show the
size of the bin.
The ratio of each MC prediction to the data is also shown in the lower part of each plot. 
}
\label{fig-Diff2}
\end{center}
\end{figure}

\begin{figure}[p]
\begin{center}							       
\includegraphics[width=0.85\hsize,clip=true,viewport=0 0 540 610] {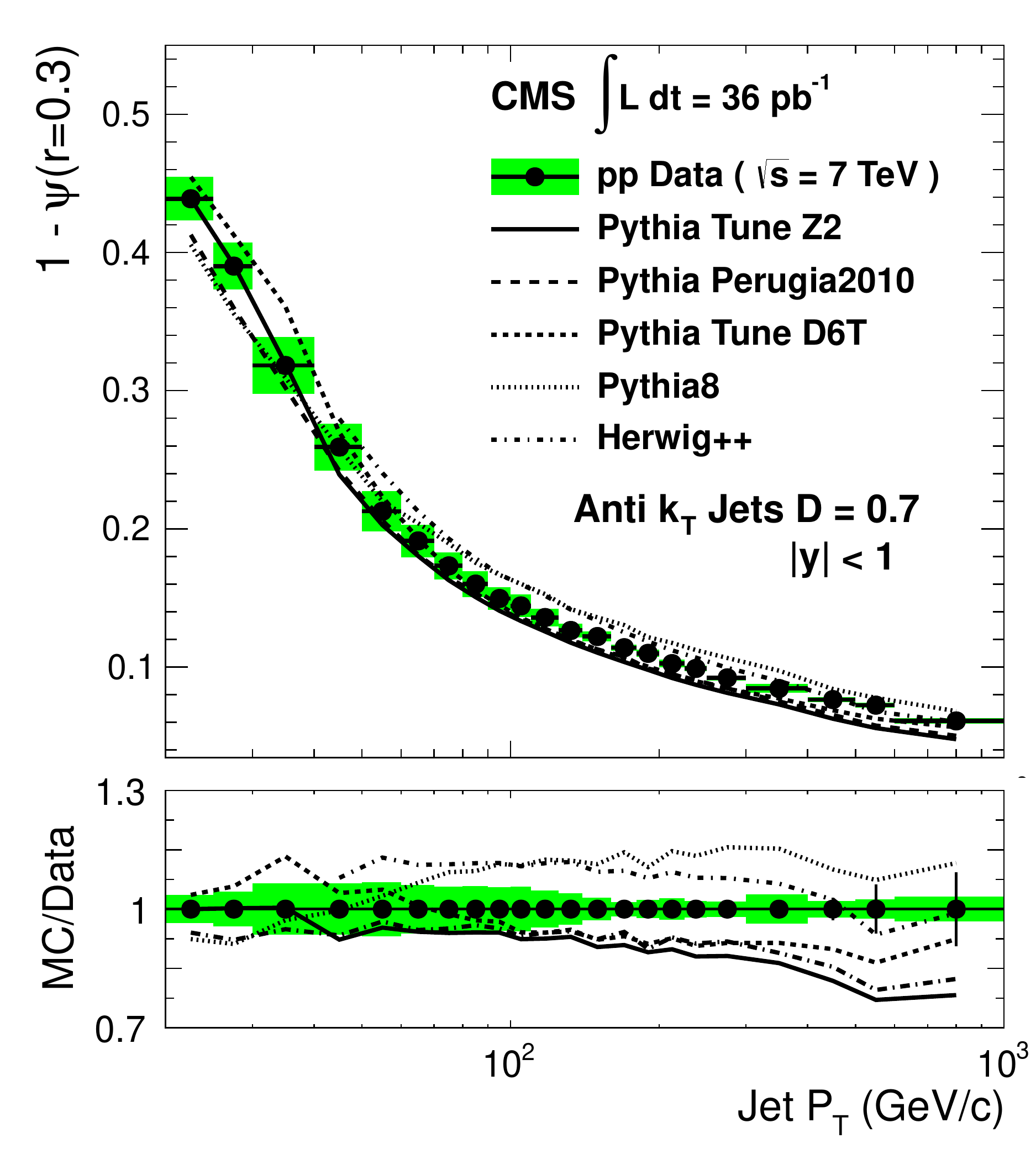}
\caption{Measured integrated jet shape, $1 - \Psi(r=0.3)$, as a function of jet $\pt$ in the central rapidity region $|y|<1$, 
compared to {\sc herwig++}, {\sc pythia8}, and {\sc pythia6} predictions with various tunes. 
Statistical uncertainties are shown as uncertainties on the data points and
the shaded region represents the total systematic uncertainty of the measurement.
Data points are placed at the bin centre; the horizontal bars show the
size of the bin.
The ratio of each MC prediction to the data is also shown in the lower part of each plot. 
}
\label{CentralJSvspt}
\end{center}
\end{figure}

\begin{figure}[p]
\begin{center}

\includegraphics[width=0.35\hsize,clip=false,viewport=0 0 550 650] {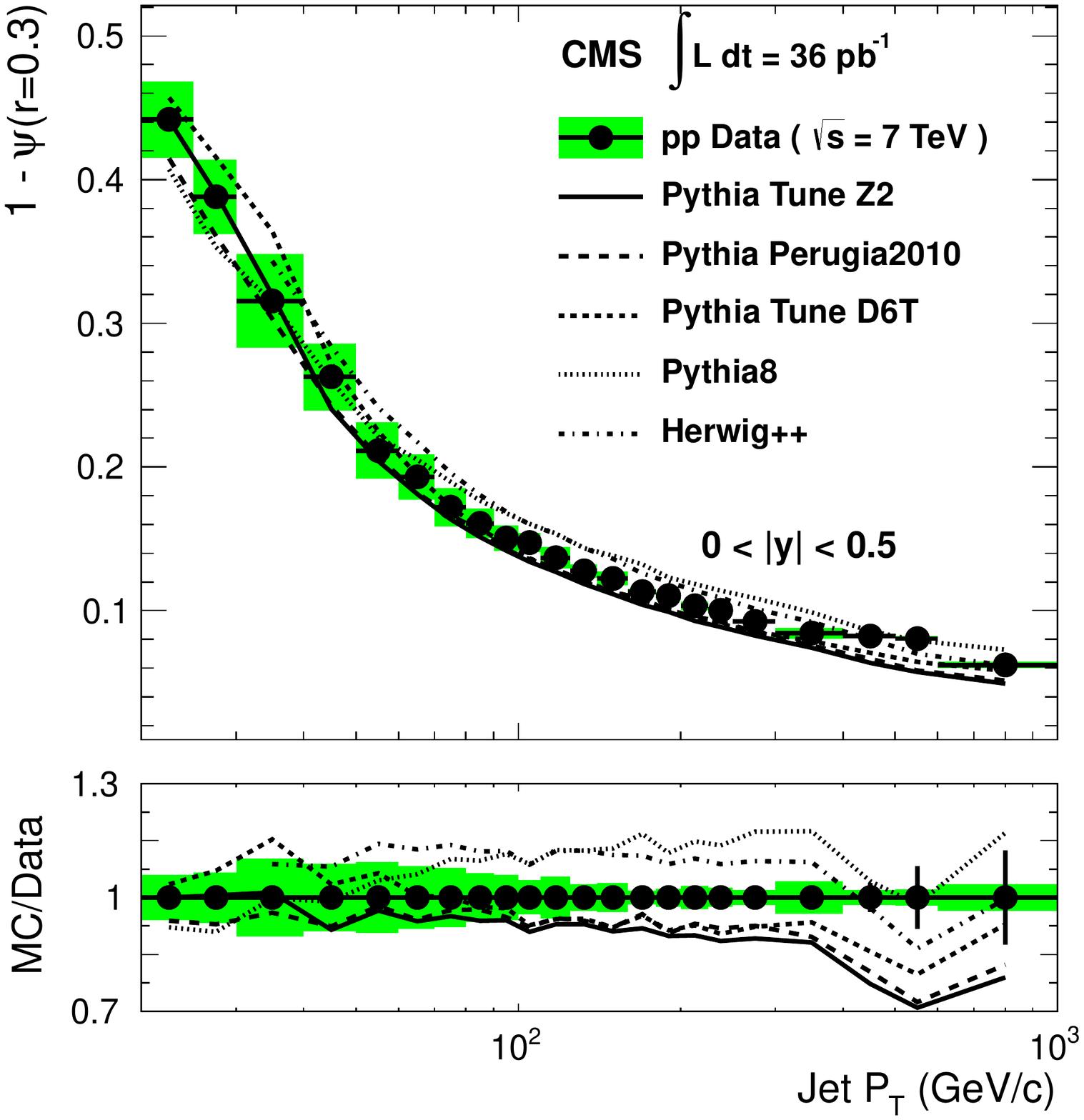}
\includegraphics[width=0.35\hsize,clip=false,viewport=0 0 550 650] {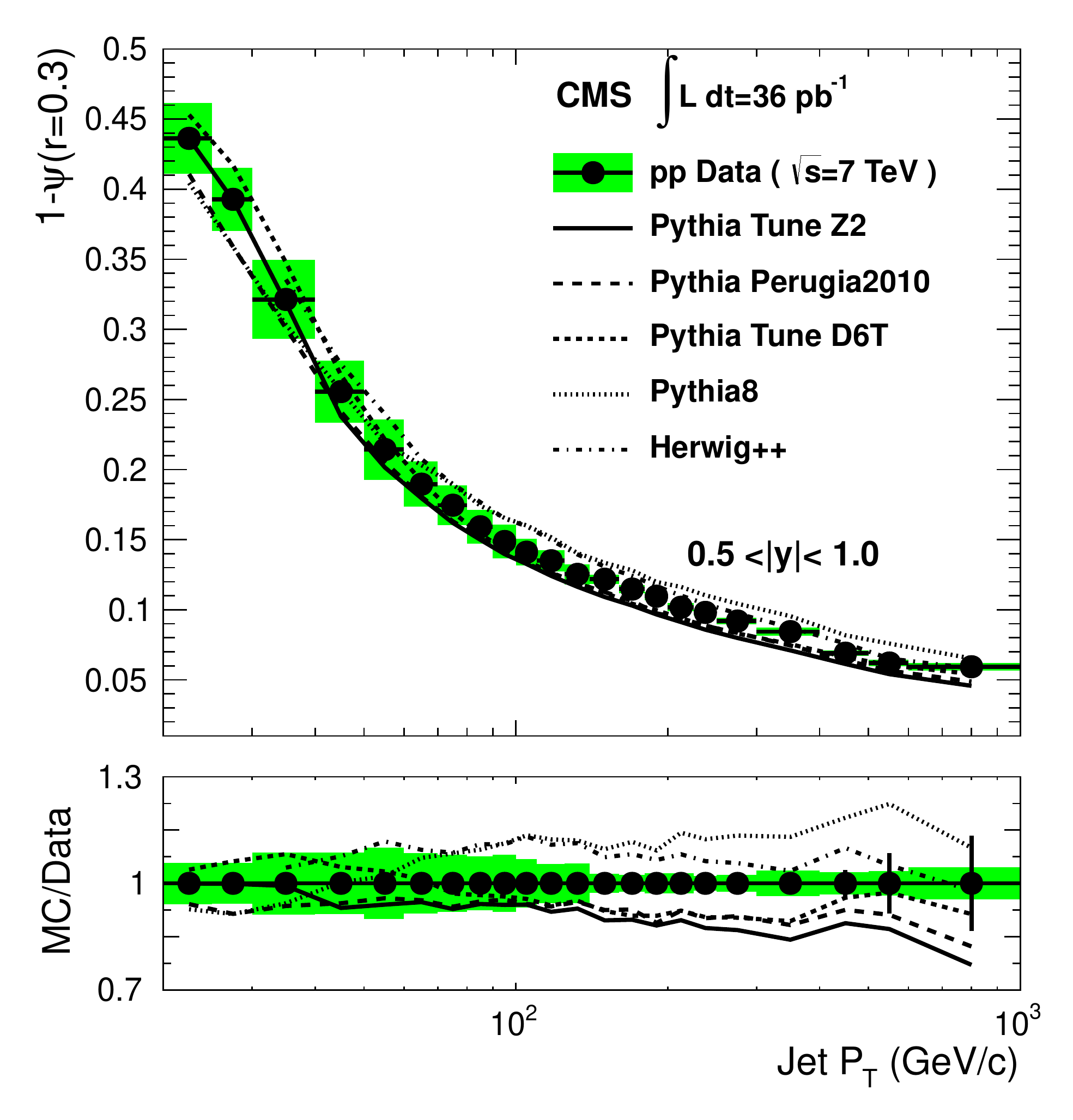}
\includegraphics[width=0.35\hsize,clip=false,viewport=0 0 550 650] {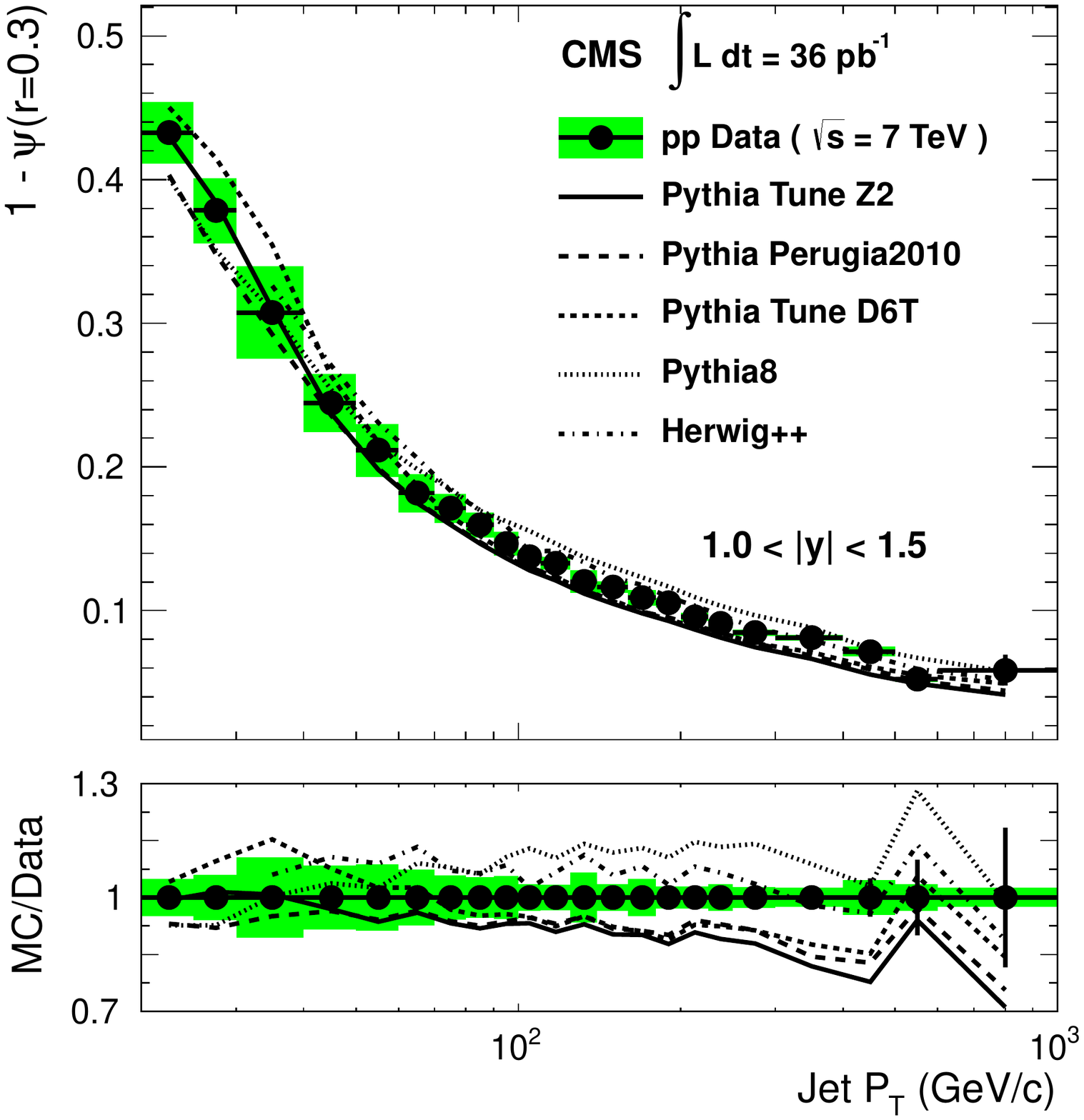}
\includegraphics[width=0.35\hsize,clip=false,viewport=0 0 550 650] {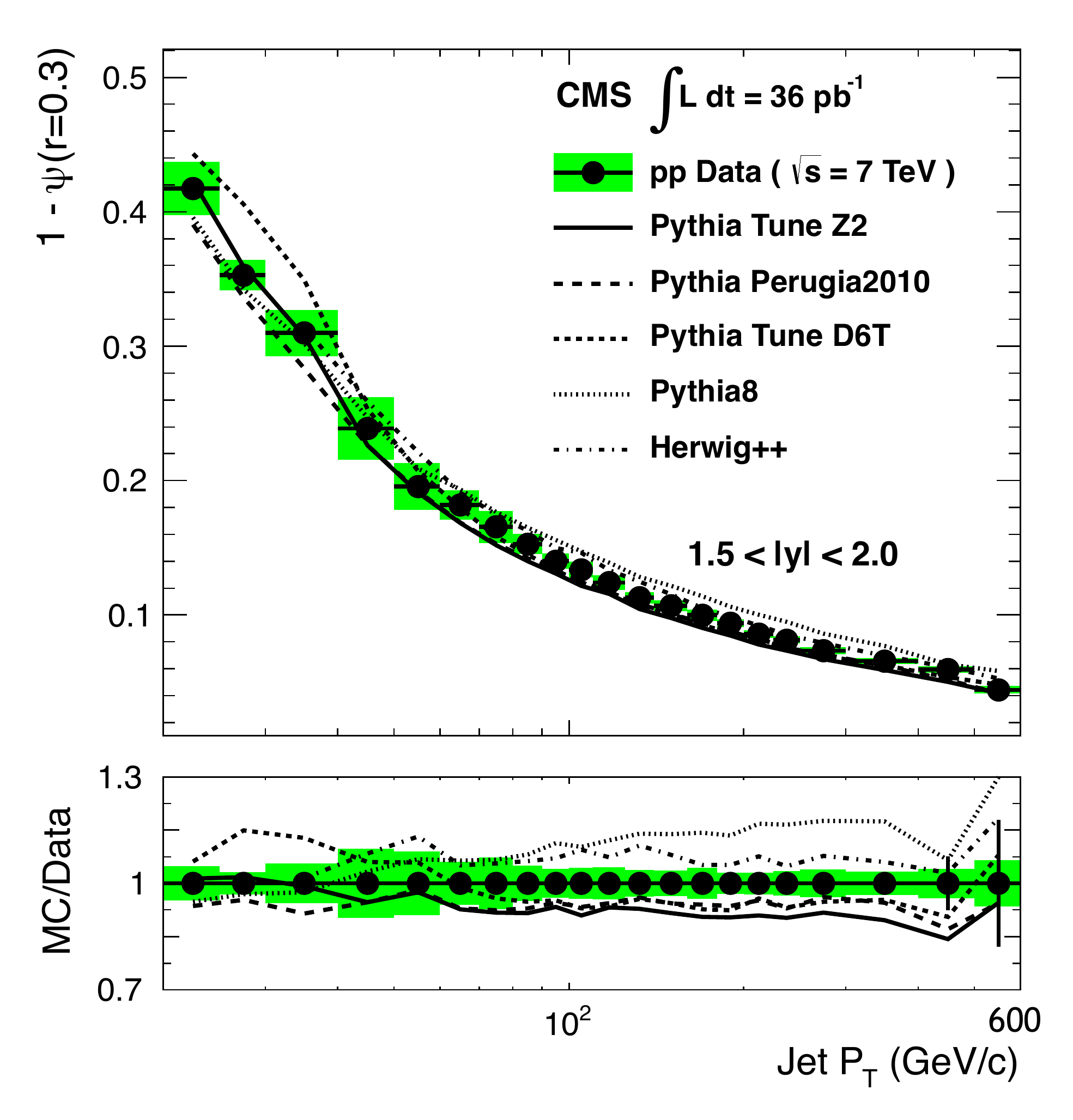}
\includegraphics[width=0.35\hsize,clip=false,viewport=0 0 550 650] {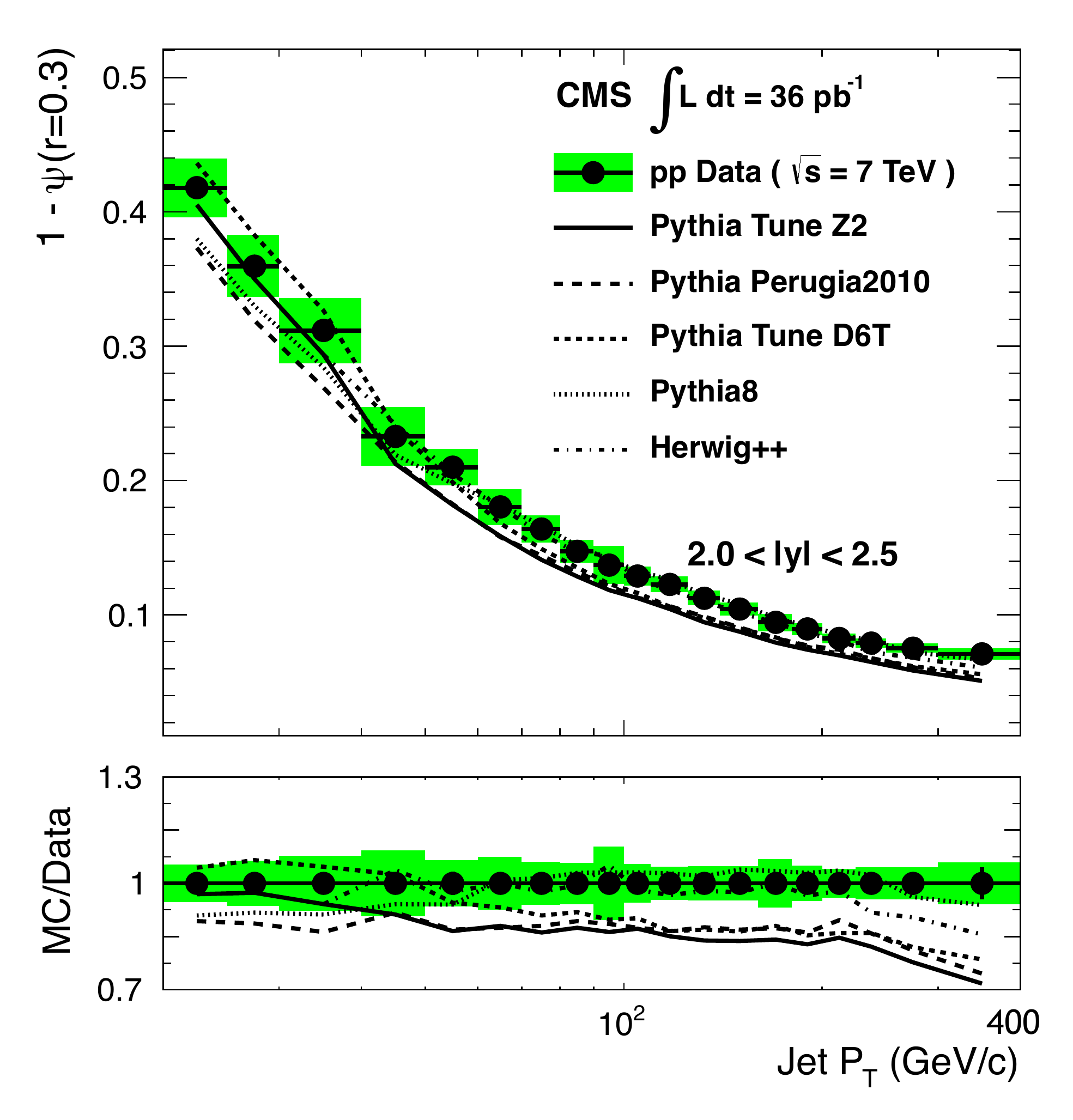}
\includegraphics[width=0.35\hsize,clip=false,viewport=0 0 550 650] {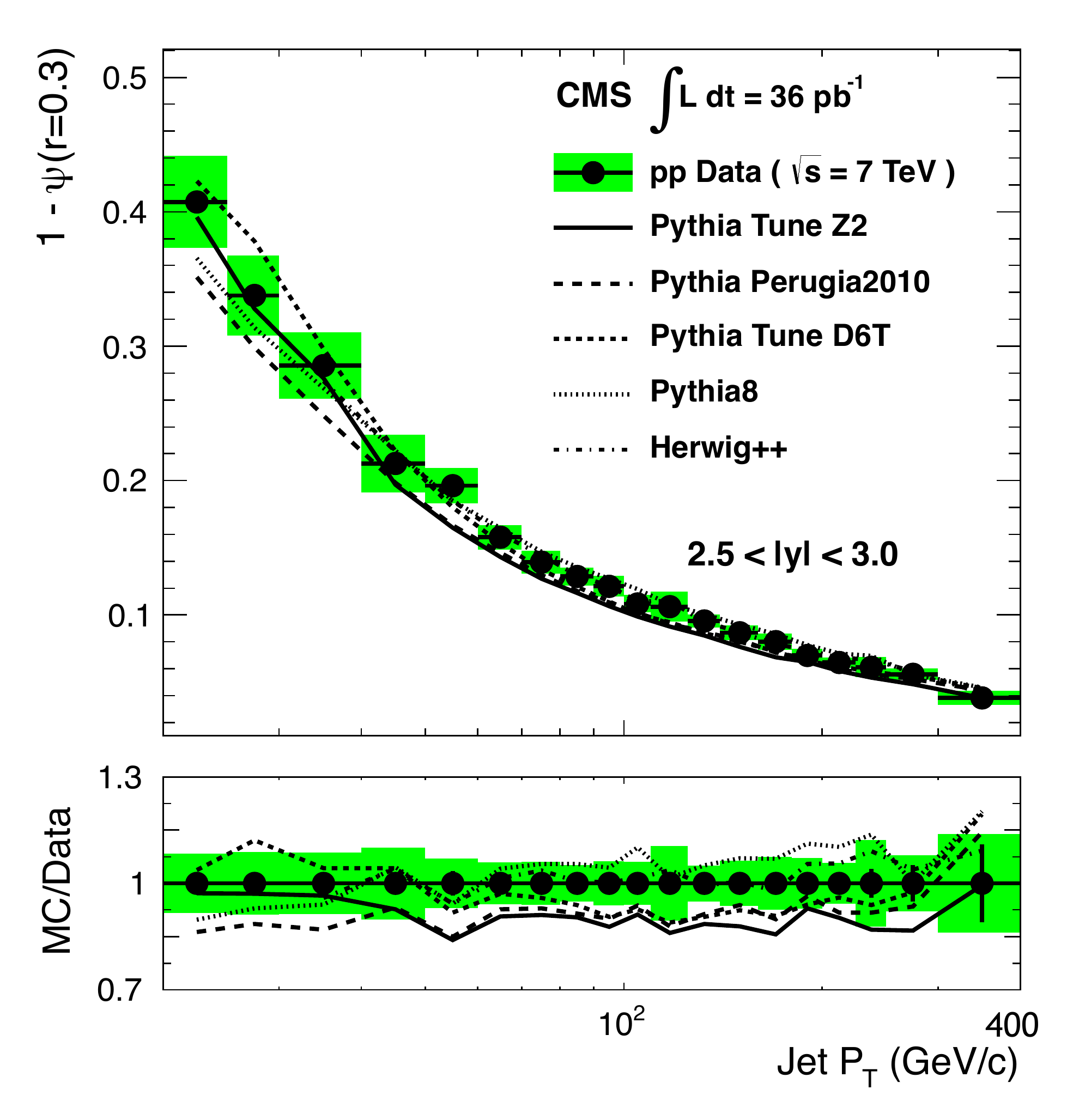}

\caption{Measured integrated jet shape, $1 - \Psi(r=0.3)$, as a function of jet $\pt$ in different 
jet rapidity regions, 
compared to {\sc herwig++}, {\sc pythia8}, and {\sc pythia6} predictions with various tunes.
Statistical uncertainties are shown as error bars on the data points and
the shaded region represents the total systematic uncertainty of the measurement.
Data points are placed at the bin centre; the horizontal bars show the
size of the bin. 
The ratio of each MC prediction to the data is also shown in the lower part of each plot. 
} 
\label{shapeVspt_eta}
\end{center}
\end{figure}

The differential jet shape measurements for central jets 
($|y|<1$) for representative bins in jet \pt, 
along with their statistical and systematic uncertainties, compared with predictions from different MC generators and tunes
are presented in Figs.~\ref{fig-Diff1} and \ref{fig-Diff2}.
Larger values of $\rho(r)$ denote larger transverse momentum fraction in a particular annulus. At high jet \pt, 
the data are peaked at low radius $r$, indicating that jets are highly collimated with most 
of their \pt close to the jet axis while they widen at lower jet \pt.
For the lowest jet \pt bins, the \pt distribution within the jet
flattens considerably. 
For 20\GeVc jets in the central rapidity region, approximately 15\% of the jet \pt is within
a radius of $r=0.1$ around the jet axis, whereas at 600\GeVc this fraction increases to about 90\%.
This behaviour is illustrated in 
Figs.~\ref{CentralJSvspt} and \ref{shapeVspt_eta} where the amount of jet energy deposited outside a cone of $r=0.3$,
$1-\Psi(r=0.3)$, is shown
as a function of jet \pt for central jets and also in six different jet rapidity regions up to $|y|=3$.
These figures also show comparisons of the data with the {\sc pythia6}, {\sc pythia8}, and {\sc herwig++} 
generators.

\begin{figure}[p]
\begin{center}
\includegraphics[width=0.49\hsize,clip=false,viewport=0 0 600 650] {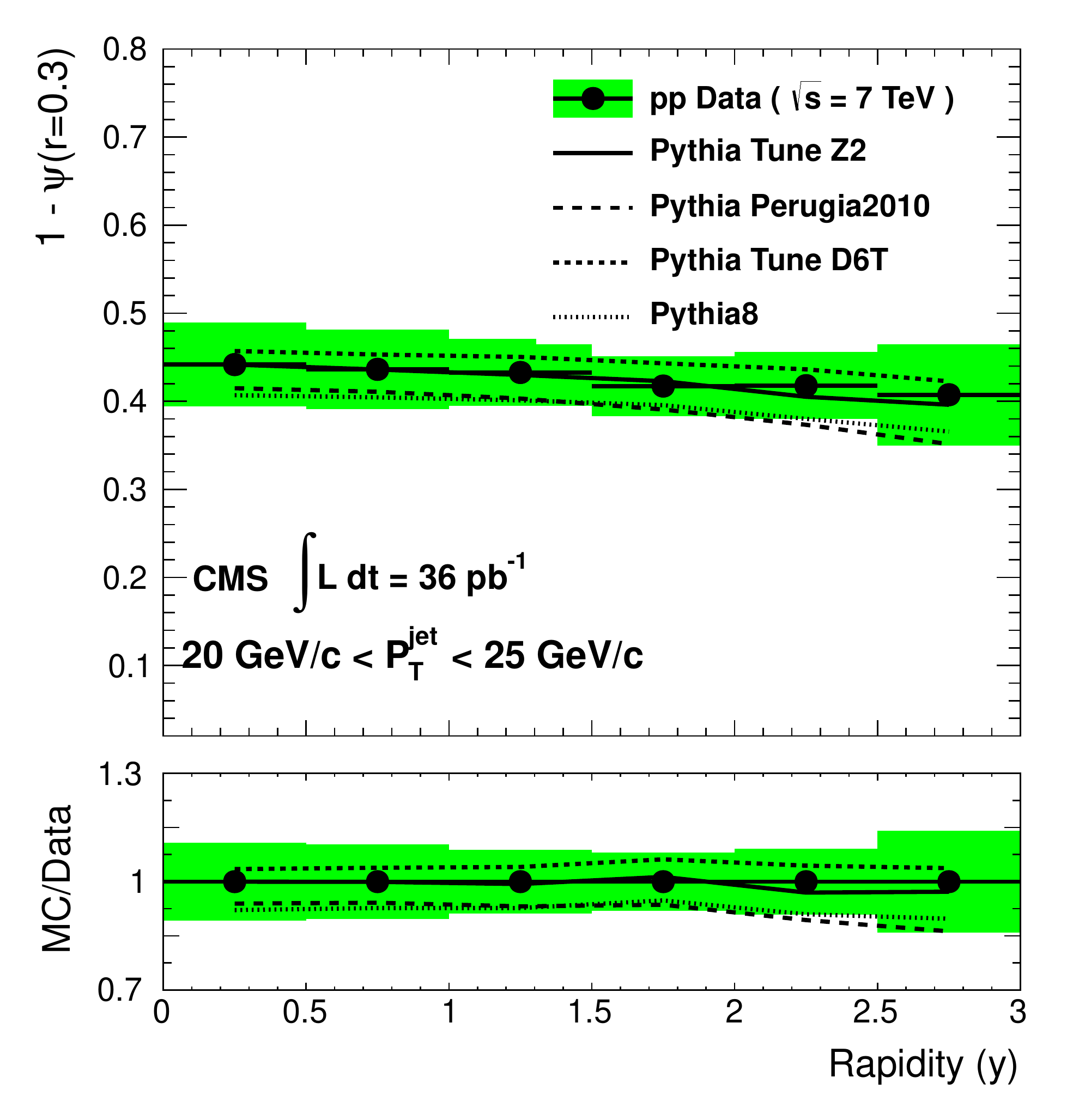}
\includegraphics[width=0.49\hsize,clip=false,viewport=0 0 600 650] {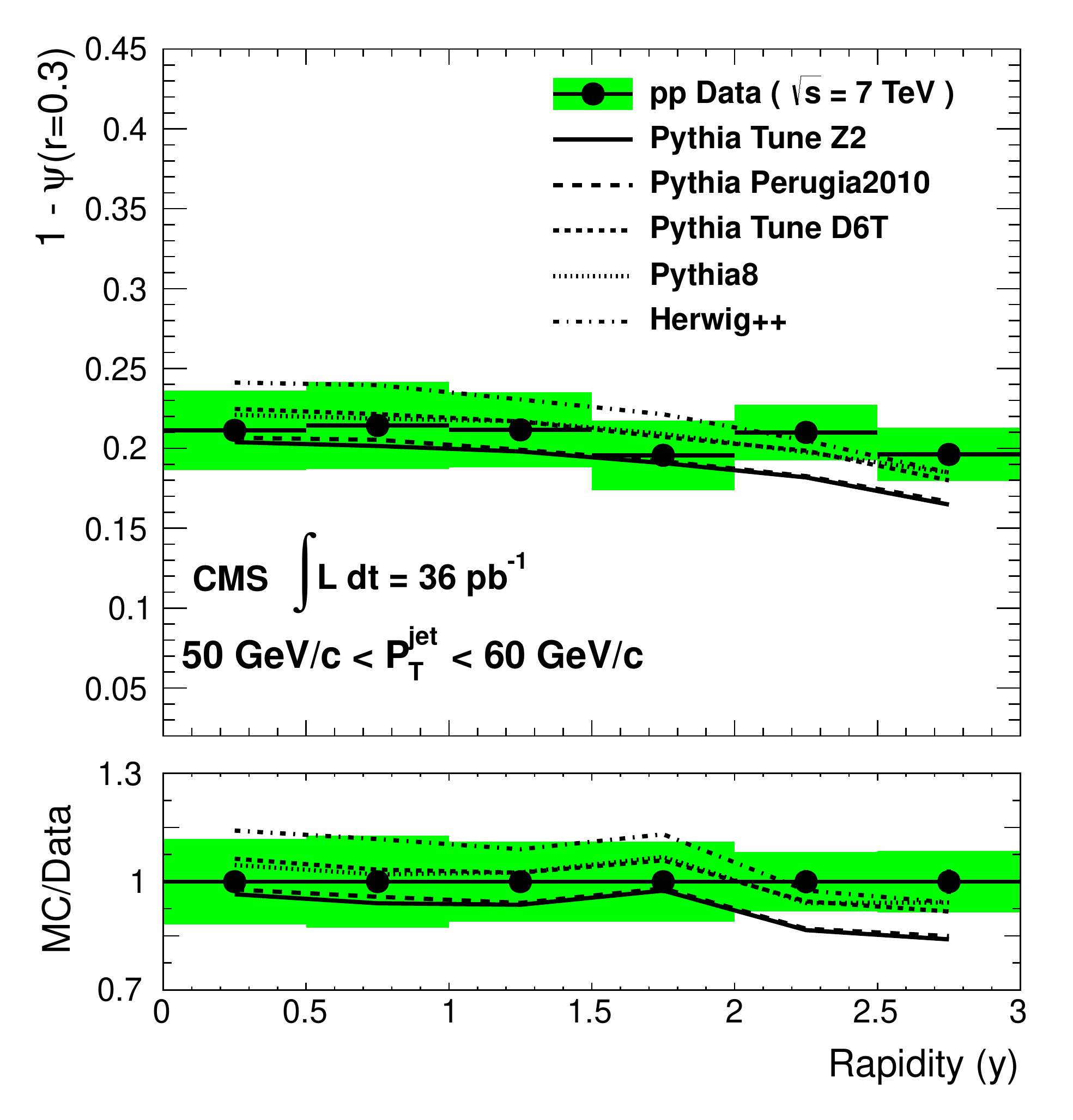}
\includegraphics[width=0.49\hsize,clip=false,viewport=0 0 600 650] {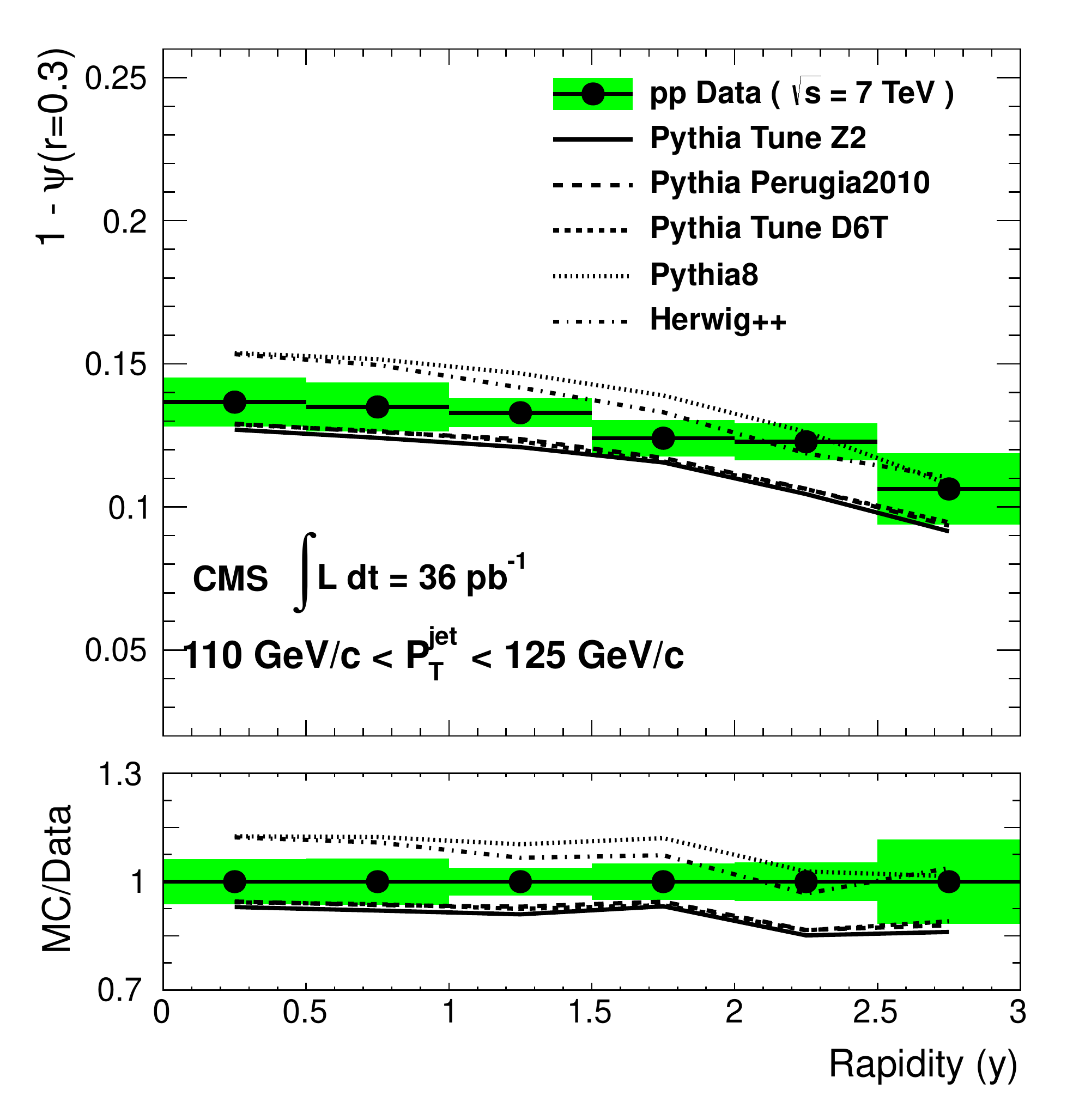}
\includegraphics[width=0.49\hsize,clip=false,viewport=0 0 600 650] {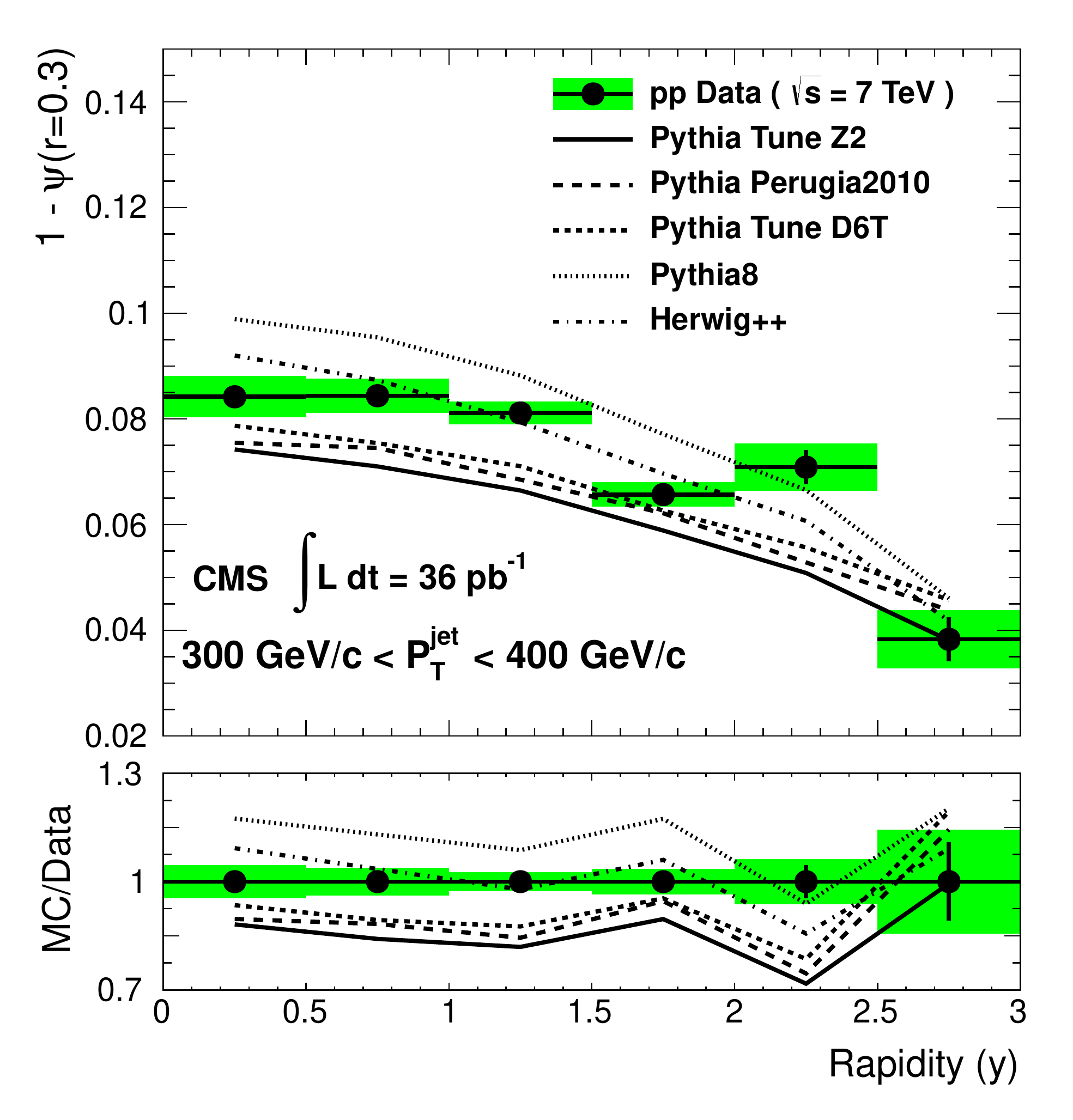}
\caption{Measured integrated jet shape, $1 - \Psi(r=0.3)$, as a function of jet rapidity for representative jet \pt bins.
The data (points) are compared to particle-level 
{\sc herwig++}, {\sc pythia8}, and {\sc pythia6} predictions with various tunes. 
Statistical uncertainties are shown as error bars on the data points and
the shaded region represents the total systematic uncertainty of the measurement.
Data points are placed at the bin centre; the horizontal bars show the
size of the bin.
The ratio of each MC prediction to the data is also shown in the lower part of each plot. 
}
\label{ShapeVsY_pT}
\end{center}
\end{figure}

As depicted in Figs.~\ref{CentralJSvspt} and \ref{shapeVspt_eta}, at low jet \pt the {\sc pythia8} generator predicts
somewhat narrower jets than those found in data, while {\sc pythia6} tune D6T predicts wider
jets. Tune Z2 provides a good description of data at low jet \pt.
At jet $\pt\gtrsim40$\GeVc the Perugia2010 and D6T tunes describe the data better than tune Z2.
This trend holds for all rapidity ranges. 
{\sc herwig++} predicts wider jets than observed in data over most of the jet \pt region except at the forward
rapidity regions where the agreement is better.
The measurement is presented as a function of jet rapidity for
different \pt regions in Fig.~\ref{ShapeVsY_pT}, which shows that jets 
become somewhat narrower with increasing $|y|$ in both data and simulation.
\begin{figure}[p]
\begin{center}
  \includegraphics[width=0.60\textwidth]{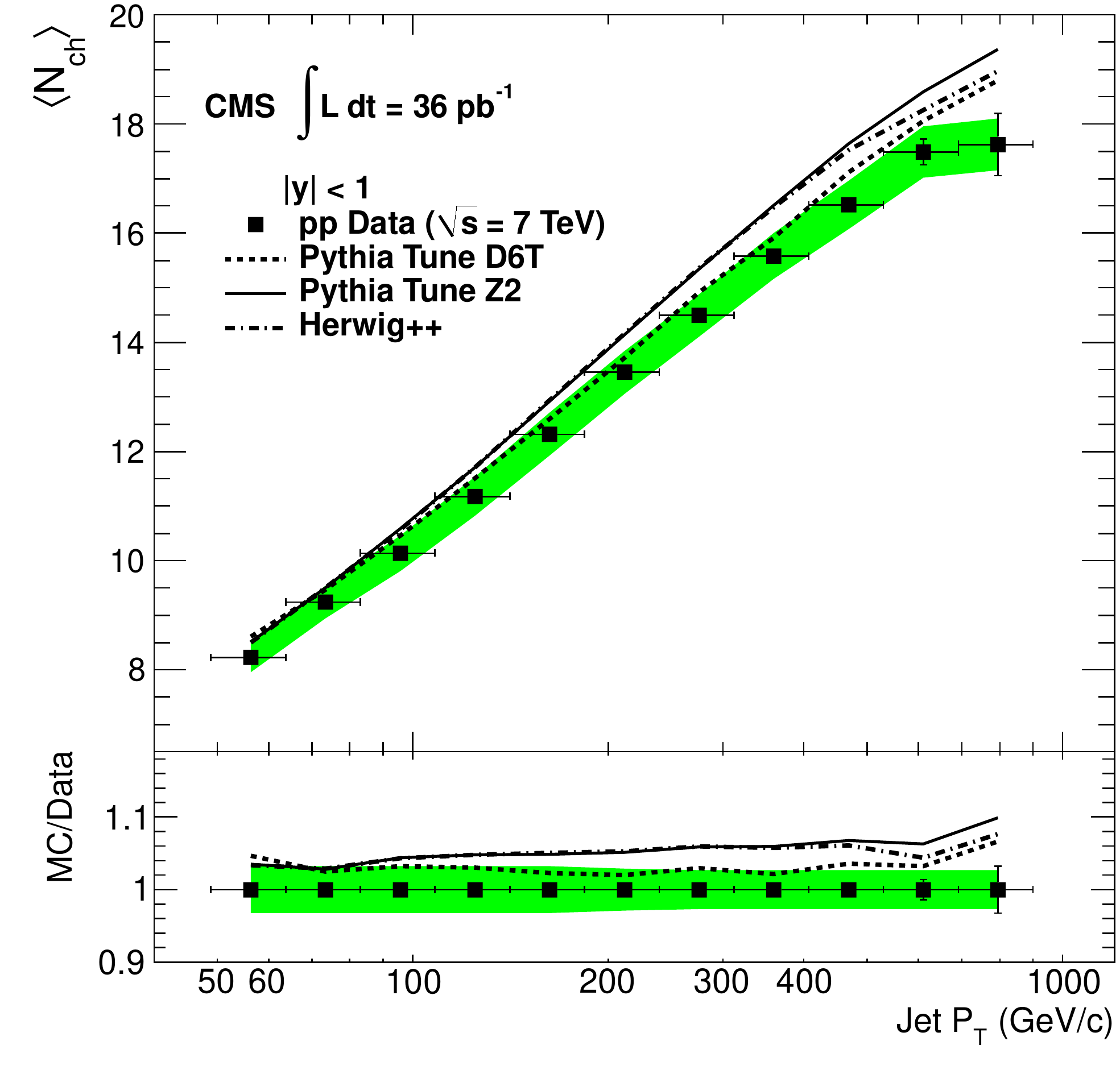}
  \includegraphics[width=0.60\textwidth]{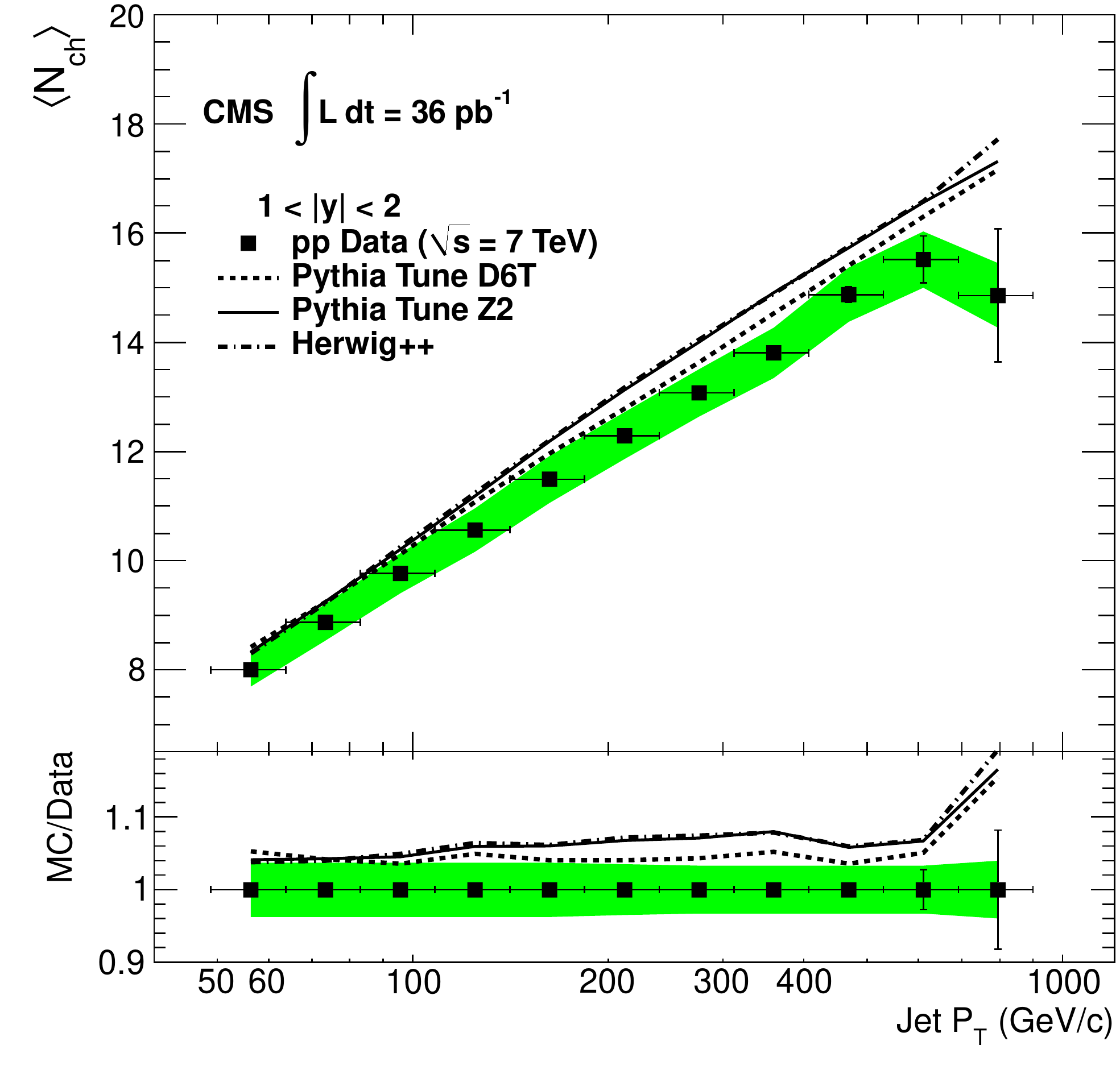}
  \caption{The average charged-particle multiplicity \Nch as a function of jet \pt for
    jets with $0<|y|<1$ (top) and $1<|y|<2$ (bottom). 
    Data are shown with statistical error bars and a band denoting the
    systematic uncertainty. 
    Also shown are predictions based on the {\sc pythia6} tune D6T (dashed
    line) and tune Z2 (solid line) and {\sc herwig++} (dot-dashed line) event generators.
    The bottom of each plot shows the ratio of the MC simulations to data with statistical error 
    bars and a band denoting the systematic uncertainties on the data measurement.}
  \label{fig:dnch_12}
\end{center}
\end{figure}
\begin{figure}[p]
\begin{center}
  \includegraphics[width=0.60\textwidth]{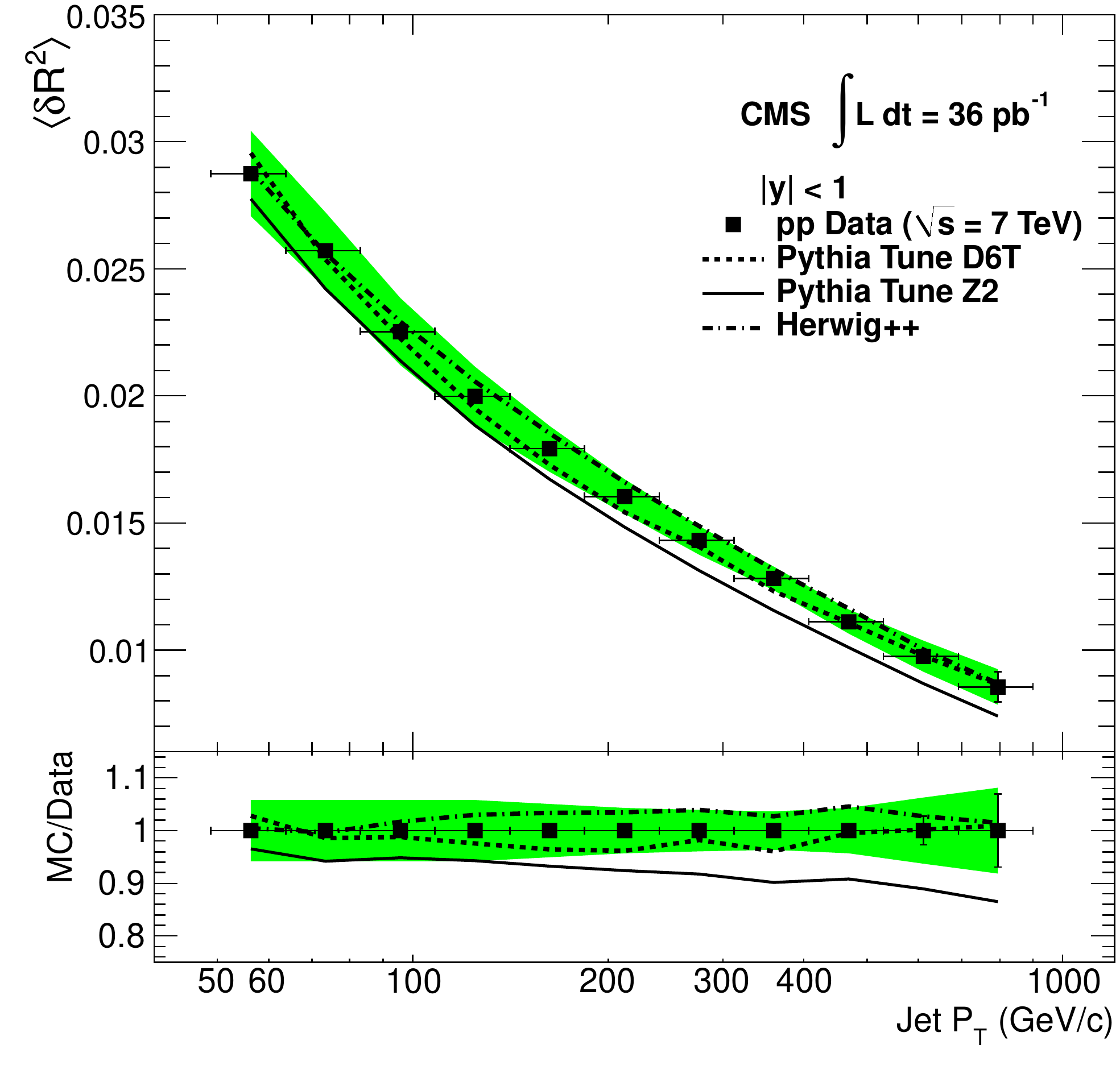}
  \includegraphics[width=0.60\textwidth]{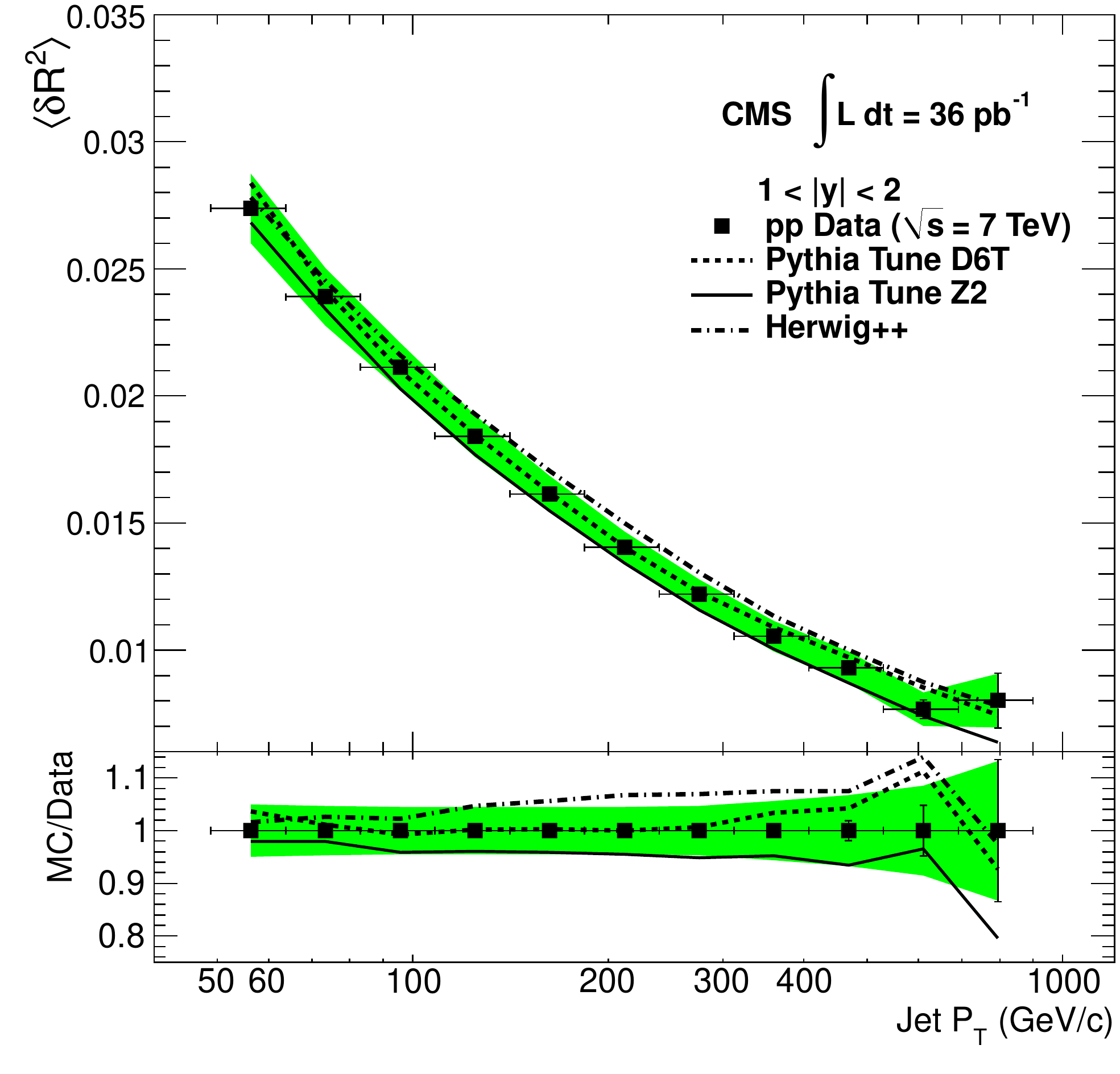}
  \caption{The average transverse jet size \deltaR as a function of jet \pt for 
    jets with $0<|y|<1$ (top) and $1<|y|<2$ (bottom).
    Data are shown with statistical error bars and a band denoting the
    systematic uncertainty. 
    Predictions are shown based on the {\sc pythia6} tune D6T (dashed
    line), tune Z2 (solid line), and {\sc herwig++} (dot-dashed line) event generators.
    The bottom of each plot shows the ratio of the MC simulations to data with statistical error 
    bars and a band denoting the systematic uncertainties on the data measurement.}
  \label{fig:dr2_12}
\end{center}
\end{figure}
The measured \Nch and \deltaR as functions of
jet \pt are presented in 
Figs.~\ref{fig:dnch_12} and \ref{fig:dr2_12} for two different rapidity intervals, $|y|<1$ and $1<|y|<2$,
along with their statistical and systematic uncertainties.
The total systematic uncertainty includes 
the uncertainty on the jet energy scale, 
jet energy resolution, tracking inefficiency, jet unsmearing
procedure, and pileup contribution. The ratios of the MC predictions to data, corrected
to the particle level, of these two observables
are shown at the bottom of the figures.
The measured values of \Nch are systematically lower than 
the values predicted by both {\sc pythia6} and {\sc herwig++}. In the case of 
\deltaR the predicted values are in agreement with the measured 
values with the exception of some disagreement observed with {\sc pythia6} tune Z2 at 
$|y|<1$.

\begin{figure}[htb]
\begin{center}
   \resizebox{10cm}{!}{\includegraphics{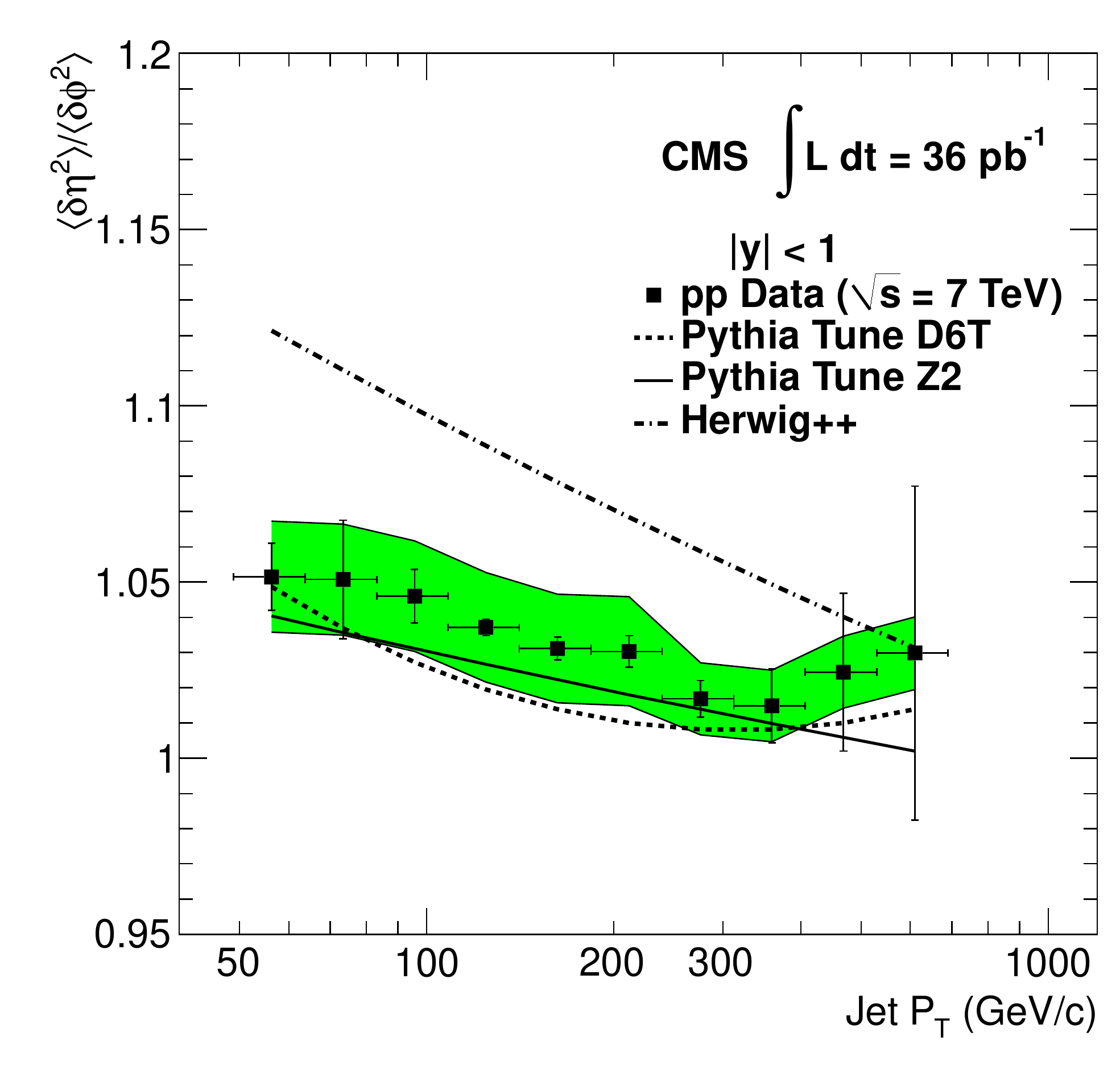}}
      \caption{The 
        ratio of the jet transverse width second moments in the $\eta$ and $\phi$
        directions as a function of jet \pt for jets with $|y|<1$. The systematic uncertainty is shown
        as a band around the data points. Also shown are predictions based on the 
        {\sc pythia6} tune D6T (dot-dashed line), tune Z2 (solid line), and {\sc herwig++} (dashed line) event generators.}
    \label{fig:detadephi_PAS}
\end{center}
\end{figure}

The ratio of the second moments in the $\eta$ and $\phi$ directions 
is shown as a function of jet \pt for $|y|<1$ in Fig.~\ref{fig:detadephi_PAS}.
Systematic uncertainties largely cancel in this ratio. 
The measured jet width in the $\eta$ direction is slightly wider than in the $\phi$ direction. 
These results agree with {\sc pythia6} predictions, while 
{\sc herwig++} predicts a larger difference of the jet width in 
the $\eta$ and $\phi$ directions. 

\begin{figure}[p]
\begin{center}
  \includegraphics[width=0.60\textwidth]{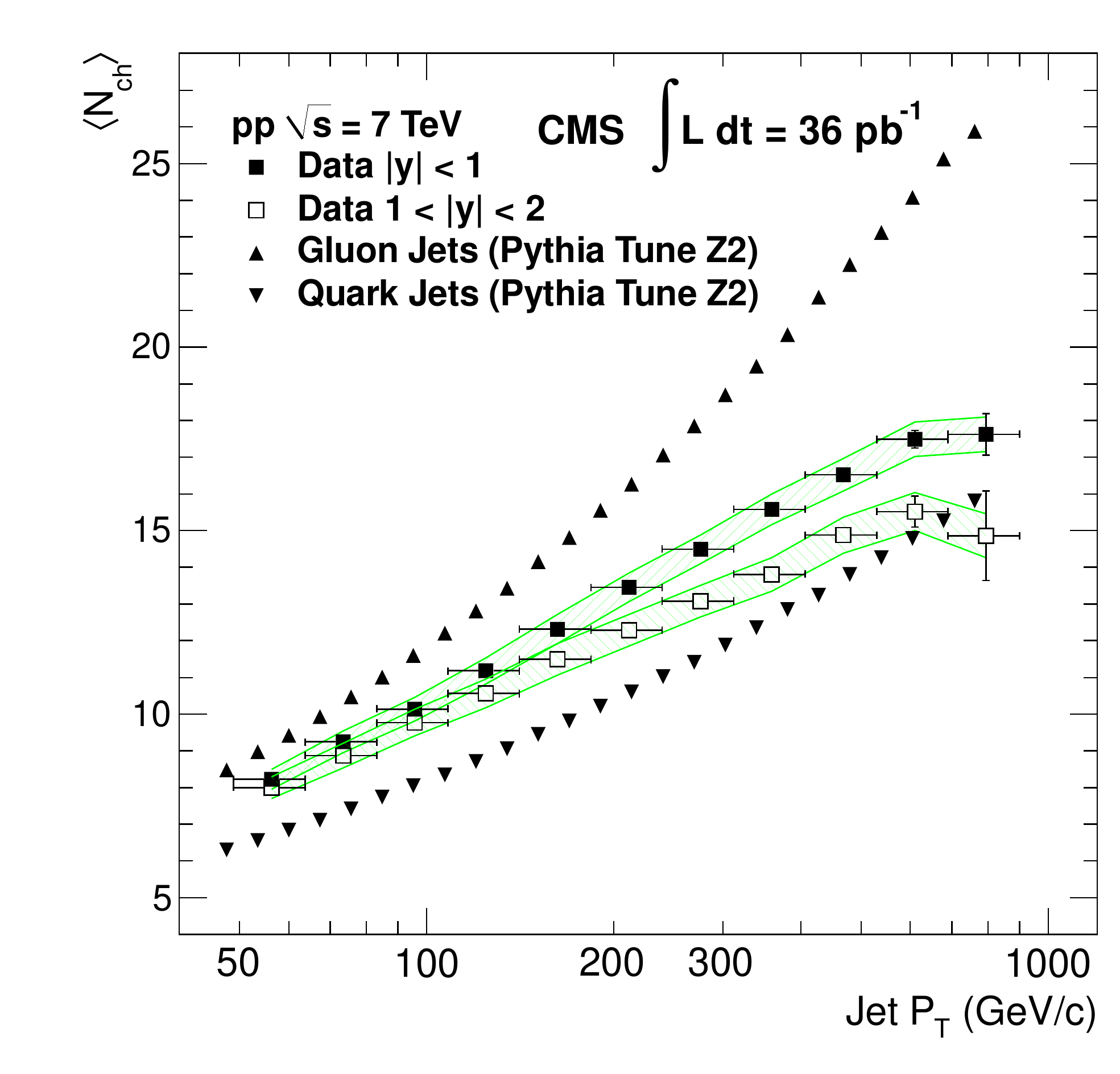}
  \includegraphics[width=0.60\textwidth]{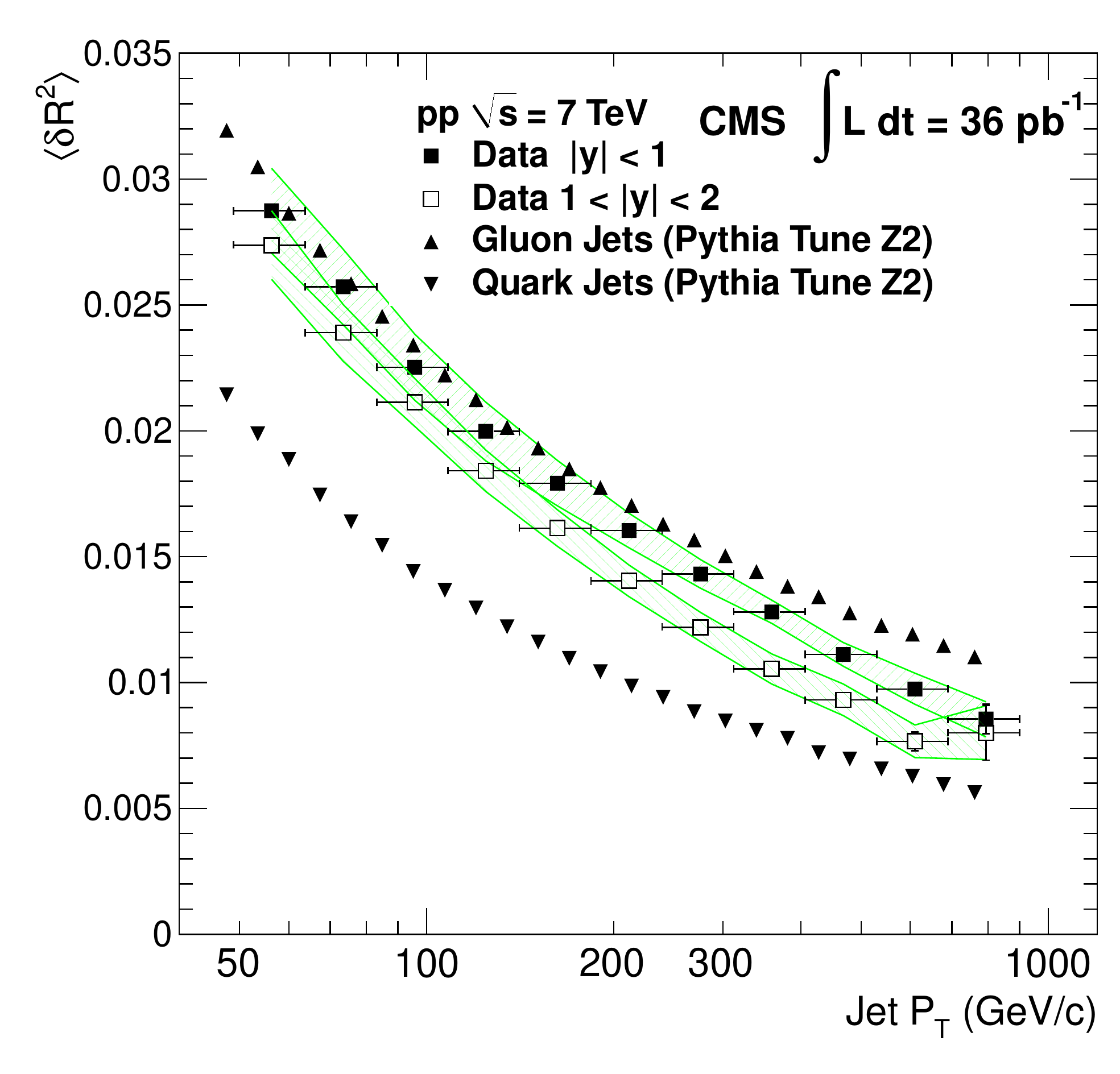}
  \caption{Average charged-particle multiplicity \Nch (top) and average transverse jet size 
    \deltaR (bottom) as
    functions of jet \pt for jets with $0 < |y| < 1$ (solid squares)
    and with $1 < |y| < 2$ (open squares). Data are shown with statistical error bars
    and a band denoting the systematic uncertainty. Also shown are predictions for
    quark-induced and gluon-induced jets for $|y| < 1$ based on the {\sc pythia6} tune Z2 event generator.}
  \label{fig:nch_dr2_12}
\end{center}
\end{figure}

A comparison of the \Nch and \deltaR values obtained from the data 
as functions of jet \pt in two ranges of jet rapidity 
is shown in Fig.~\ref{fig:nch_dr2_12}. 
The data are in good agreement with the hypothesis that the fraction of quark-induced jets 
increases with increasing jet \pt and jet rapidity.

Tables containing the measured jet shape, charged-hadron multiplicity, and transverse size data are available as a 
supplement to the online version of this article.

\section {Summary}
We have presented measurements of jet shapes, mean charged-hadron multiplicity,
and transverse width for jets produced in proton-proton collisions at a centre-of-mass 
energy of 7\TeV, collected by the CMS detector at the LHC. 
Jets become narrower with increasing jet \pt, 
and they also show a mild rapidity dependence in which jets become somewhat narrower with increasing $|y|$,
in the manner predicted by various QCD Monte Carlo models.
At low jet \pt, the {\sc pythia6} Z2 model tuned to the initial CMS soft \pt data~\cite{UE_paper}
provides a fair description of the measured jet shapes. At jet $\pt\gtrsim40$\GeVc,
the tune Z2 predicts slightly narrower jets than those observed in data whereas
the D6T and Perugia2010 tunes describe the data better.
The measurements may be used to further improve these Monte Carlo models.

The mean charged-hadron multiplicity and the second
moment of the jet width are 
compared with predictions from 
the {\sc pythia6} (tunes D6T and Z2) and {\sc herwig++} generators. 
All these models predict 
slightly higher mean charged-hadron multiplicities than found in the 
data; however, good agreement is observed between the models and the measured second moment of the
jet transverse width.
The observed behaviour of the mean multiplicity and 
jet transverse width agrees with the predicted 
increase in the fraction of quark-induced
jets at higher jet transverse 
momentum and rapidity. Decomposition of the transverse width second moment into 
second moments for $\eta$ and $\phi$ demonstrates that 
jets are slightly wider in the $\eta$ direction than in the $\phi$ direction. This observation 
is in good quantitative agreement with {\sc pythia6} predictions,
while {\sc herwig++} predicts a larger difference between jet widths in the $\eta$ and 
$\phi$ directions. 

\section*{Acknowledgements}

\hyphenation{Bundes-ministerium Forschungs-gemeinschaft Forschungs-zentren} We congratulate our colleagues in the CERN accelerator departments for the excellent performance of the LHC machine. We thank the technical and administrative staff at CERN and other CMS institutes. This work was supported by the Austrian Federal Ministry of Science and Research; the Belgium Fonds de la Recherche Scientifique, and Fonds voor Wetenschappelijk Onderzoek; the Brazilian Funding Agencies (CNPq, CAPES, FAPERJ, and FAPESP); the Bulgarian Ministry of Education and Science; CERN; the Chinese Academy of Sciences, Ministry of Science and Technology, and National Natural Science Foundation of China; the Colombian Funding Agency (COLCIENCIAS); the Croatian Ministry of Science, Education and Sport; the Research Promotion Foundation, Cyprus; the Ministry of Education and Research, Recurrent financing contract SF0690030s09 and European Regional Development Fund, Estonia; the Academy of Finland, Finnish Ministry of Education and Culture, and Helsinki Institute of Physics; the Institut National de Physique Nucl\'eaire et de Physique des Particules~/~CNRS, and Commissariat \`a l'\'Energie Atomique et aux \'Energies Alternatives~/~CEA, France; the Bundesministerium f\"ur Bildung und Forschung, Deutsche Forschungsgemeinschaft, and Helmholtz-Gemeinschaft Deutscher Forschungszentren, Germany; the General Secretariat for Research and Technology, Greece; the National Scientific Research Foundation, and National Office for Research and Technology, Hungary; the Department of Atomic Energy and the Department of Science and Technology, India; the Institute for Studies in Theoretical Physics and Mathematics, Iran; the Science Foundation, Ireland; the Istituto Nazionale di Fisica Nucleare, Italy; the Korean Ministry of Education, Science and Technology and the World Class University program of NRF, Korea; the Lithuanian Academy of Sciences; the Mexican Funding Agencies (CINVESTAV, CONACYT, SEP, and UASLP-FAI); the Ministry of Science and Innovation, New Zealand; the Pakistan Atomic Energy Commission; the Ministry of Science and Higher Education and the National Science Centre, Poland; the Funda\c{c}\~ao para a Ci\^encia e a Tecnologia, Portugal; JINR (Armenia, Belarus, Georgia, Ukraine, Uzbekistan); the Ministry of Education and Science of the Russian Federation, the Federal Agency of Atomic Energy of the Russian Federation, Russian Academy of Sciences, and the Russian Foundation for Basic Research; the Ministry of Science and Technological Development of Serbia; the Ministerio de Ciencia e Innovaci\'on, and Programa Consolider-Ingenio 2010, Spain; the Swiss Funding Agencies (ETH Board, ETH Zurich, PSI, SNF, UniZH, Canton Zurich, and SER); the National Science Council, Taipei; the Scientific and Technical Research Council of Turkey, and Turkish Atomic Energy Authority; the Science and Technology Facilities Council, UK; the US Department of Energy, and the US National Science Foundation.

Individuals have received support from the Marie-Curie programme and the European Research Council (European Union); the Leventis Foundation; the A. P. Sloan Foundation; the Alexander von Humboldt Foundation; the Belgian Federal Science Policy Office; the Fonds pour la Formation \`a la Recherche dans l'Industrie et dans l'Agriculture (FRIA-Belgium); the Agentschap voor Innovatie door Wetenschap en Technologie (IWT-Belgium); the Council of Science and Industrial Research, India; and the HOMING PLUS programme of Foundation for Polish Science, cofinanced from European Union, Regional Development Fund.

\bibliography{auto_generated}   

\cleardoublepage \appendix\section{The CMS Collaboration \label{app:collab}}\begin{sloppypar}\hyphenpenalty=5000\widowpenalty=500\clubpenalty=5000\textbf{Yerevan Physics Institute,  Yerevan,  Armenia}\\*[0pt]
S.~Chatrchyan, V.~Khachatryan, A.M.~Sirunyan, A.~Tumasyan
\vskip\cmsinstskip
\textbf{Institut f\"{u}r Hochenergiephysik der OeAW,  Wien,  Austria}\\*[0pt]
W.~Adam, T.~Bergauer, M.~Dragicevic, J.~Er\"{o}, C.~Fabjan, M.~Friedl, R.~Fr\"{u}hwirth, V.M.~Ghete, J.~Hammer\cmsAuthorMark{1}, M.~Hoch, N.~H\"{o}rmann, J.~Hrubec, M.~Jeitler, W.~Kiesenhofer, M.~Krammer, D.~Liko, I.~Mikulec, M.~Pernicka$^{\textrm{\dag}}$, B.~Rahbaran, C.~Rohringer, H.~Rohringer, R.~Sch\"{o}fbeck, J.~Strauss, A.~Taurok, F.~Teischinger, P.~Wagner, W.~Waltenberger, G.~Walzel, E.~Widl, C.-E.~Wulz
% ...
\vskip\cmsinstskip
\dag:~Deceased\\
1:~~Also at CERN, European Organization for Nuclear Research, Geneva, Switzerland\\
% ...

\end{sloppypar}
\end{document}